\documentclass[aps,twocolumn,amsmath,amssymb,amsfonts,floatfix,superscriptaddress]{revtex4-2}
\usepackage{epsfig}
\usepackage{epstopdf}
\usepackage{ifpdf} 
\usepackage{graphicx}
\usepackage{color}
\usepackage[usenames,dvipsnames]{xcolor}
\usepackage{bm} 
\usepackage{hyperref}
\usepackage[normalem]{ulem}
\usepackage[english]{babel}

\usepackage{natbib}
\usepackage{bbold}
\usepackage{verbatim}
\usepackage{floatrow}
\usepackage[caption=false]{subfig}
\begin{document}

	\title{Active osmotic-like pressure on permeable inclusions}

	\author{Mahmoud Sebtosheikh}
	\thanks{mahmoud-sebtosheikh@ipm.ir (corresponding author)}
\affiliation{School of Nano Science, Institute for Research in Fundamental Sciences (IPM), Tehran 19538-33511, Iran}
\affiliation{School of Physics, Institute for Research in Fundamental Sciences (IPM), Tehran 19538-33511, Iran}
\author{Ali Naji}
\affiliation{School of Nano Science, Institute for Research in Fundamental Sciences (IPM), Tehran 19538-33511, Iran}
\affiliation{Department of Physics, College of Science, Sultan Qaboos University, Muscat 123, Oman}
	\begin{abstract}
		We use a standard minimal active Brownian model to investigate the osmotic-like effective pressure generated by active fluids on fixed hollow inclusions. These inclusions are enclosed by a  permeable (albeit nonflexible) membrane, and the interior and exterior regions of the inclusions have different particle motility strengths. We consider both rectangular and disklike inclusions and analyze the effects of various system parameters, such as excluded volume interaction between active particles, hardness of membrane and active particle density, on the effective pressure produced on the enclosing membrane. We focus on the range of intermediate to high motility strengths and analyze the effective pressure in the steady state. Our findings for the active pressure produced in the interior and exterior regions of the inclusion indicate that the pressure is higher in the region with lower motility due to the relatively stronger accumulation of active particles.

	\end{abstract}
	
	\maketitle
	

	\section{Introduction}
	\label{sec:intro}

Active matter has evolved into a rapidly growing and diverse field of interest at the common interface of soft matter, statistical physics, biological sciences, and engineering \cite{revGompper2015, revLowen2016, revMarchetti2013, revPeruani2019, revStark2016, revVicsek2012, revFischer2018, revSitti2017}. Active systems comprise a wide range of biological organisms, from macroscale entities such as schools of fish and flocks of birds, to microscale entities such as bacteria, algae, and spermatozoa \cite{revMarchetti2013, revGompper2015}. These organisms have inspired the development and synthesis of many artificially made active self-propulsive particles, such as Janus colloids, photoactivated particles, and bimetallic nanorods, in the recent past \cite{revGompper2015, Paxton2004, Howse2010, Bechinger2016, Gohy2017}. Active particles continuously take up free energy from their surrounding medium and convert it to motion through internal mechanisms, such as extracellular organelles in flagellated bacteria, or chemical surface reactions, as in Janus colloids. Thanks to their out-of-equilibrium character, self-propelled particles engender intriguing many-body effects, including self-organized collective motion (see, e.g., Refs. \cite{Abkenar2013, revPeruani2019, revVicsek2012, Ramaswamy2014} and references therein), nonequilibrium clustering and phase separation \cite{Filion 2016, Graaf 2016, Dunkel 2014,Glaser2021,Abkenar2013,Dolai2020,Cates2015,Cates2018,Baskaran2013,Brady2015}, as well as long-range bath-mediated interactions between inclusions and boundary walls immersed in active suspensions \cite{Harder2014, Naji2017, Lowen2015, Cacciuto2014, Bolhuis2015, Leonardo2011, Pagonabarraga2019, Selinger2018, Naji2020z, Yang2020, Naji2020, Naji2021a, Naji2021b, Naji2021c}.

A particularly interesting facet of active suspensions is the so-called active or swim pressure produced by the constituent self-propelled particles \cite{Lowen2015,Kardar2015-1,Kardar2015-2,Cacciuto2014,Brady2014,Brady2015,Selinger2018,Marconi2017,Naji2021a,Naji2018}. In equilibrium systems, pressure can be calculated using thermodynamic, mechanical, and hydrodynamical approaches, leading to the same result. This result follows a state equation and thus varies only with bulk properties such as temperature and density. In active systems, a state equation may not generally exist \cite{Kardar2015-2}. Therefore, the pressure is mainly defined via mechanical and hydrodynamical approaches \cite{Kardar2015-1,Kardar2015-2,Brady2014}. More specifically, in the case of self-propelled spheres next to flat walls, the pressure can be described as a state function using activity-dependent effective temperature and bulk number density of spheres \cite{Brady2014,Cacciuto2014}. Active elongated particles and rods present a different situation where the pressure becomes dependent on particle-wall interactions \cite{Kardar2015-2}. Active pressure has been investigated on boundaries with different geometries such as flat walls \cite{Cacciuto2014,Selinger2018}, curved surfaces \cite{Lowen2015,Naji2021a,Naji2018,Selinger2018}, corners \cite{Lowen2015}, and sinusoidal and flexible interfaces \cite{Chen2017,ten Wolde2019}. In addition to the geometry of boundaries, active pressure can vary depending on intrinsic features of active particles such as chirality \cite{Naji2018}, interparticle interactions, and local concentration \cite{Naji2021a,Brady2014,Brady2015}.

Recently, active pressure has also been used to explain other phenomena, including motility-induced phase separation \cite{Cates2015,Cates2018,Baskaran2013,Brady2015}, active depletion \cite{Lowen2015,Selinger2018}, deformation of flexible vesicles \cite{Chen2019,Angelan2016,Quillen2020,Vutukuri2020,ten Wolde2019,Chen2017,Hagan2021}, anomalous droplet ripening \cite{Naji2018,Cates2017,Julicher2014,Hyman2014,Julicher2016}, and negative surface tension \cite{Brady2020,Lowen2015-1,Marchetti2018}. Negative surface tension emerges at the interface between dilute and condensed phases of repulsive active particles that undergo motility-induced phase separation. Despite negative surface tension, the interface is stiff and stable \cite{Marchetti2018}. Droplet ripening in active fluids transpires in a way that contrasts the so-called Ostwald ripening in equilibrium (passive) emulsions. For two interconnected droplets, while Ostwald ripening \cite{de Gennes2004} implies shrinkage of smaller droplets at the expense of the larger ones, the reversed process can take place for droplets suspended in an active fluid \cite{Naji2018,Julicher2014,Julicher2016}. In the latter case, the internal droplet pressure shows a nonmonotonic dependence on its size, enabling two interconnected droplets to reach a final state of equal size \cite{Naji2018}.
	
	The penetration of active particles through flexible membranes has been studied among other problems in the recent past. The effects of the size, shape, and activity strength of active particles, as well as the stiffness of the membrane on particle penetration, have been explored \cite{Daddi2019a, Daddi2019b, Yang2010}. These studies have identified three scenarios for penetration: trapping (active particles not being able to go through the membrane), penetration with self-healing of the membrane, and penetration with permanent disruption of the membrane \cite{Daddi2019a, Daddi2019b}. Despite the insightful results, other aspects of the problem, such as the active pressure exerted on permeable boundaries, have remained unaddressed.
	
	In this paper, we investigate the effective (osmotic-like) pressure exerted on the enclosing membrane of a permeable hollow inclusion in a suspension that involves active Brownian particles. We ignore membrane undulations and concentrate on the motility field heterogeneity, which is modeled by taking mismatching activity strengths (self-propulsion velocities) for active particles inside and outside the inclusion. The system is studied using Brownian dynamics simulation in a standard two-dimensional (2D) setting by considering both rectangular and disklike inclusions. The general aspects of the model are similar to our previous work in which effective interactions between two permeable inclusions were studied \cite{Naji2020}. In the present work, we address the behavior of effective pressure (resulting from active pressure inside and outside the inclusion) across the parameter space spanned by the inside and outside P\'eclet numbers of active particles. We also explore the dependence of effective pressure on the size and geometry of the inclusion, the hardness of its enclosing membrane, as well as interparticle interactions and the concentration of active particles.

	The model inclusions considered here can mimic soft biological micro-compartments such as cells and soft tissues, with an example being furnished by tumors invaded by active drug-delivery agents \cite{Peng2018, Joseph2017, Magdanz2019, Ghosh2020, Felfoul2016}. Artificial examples of such inclusions include fluid enclosures such as vesicles, stabilized (immiscible) droplets in emulsions, and chemically active droplets \cite{Rideau2018, Cates2017, Hyman2014, Naji2018, Vutukuri2020, Angelan2016, Kamat2011}. Spatially inhomogeneous motility fields, as considered, can also be realized by producing non-uniform temperature fields, fluid viscosity, or a heterogeneous landscape of environmental stimuli such as light and chemical reactant  \cite{Bechinger2016, Paxton2004}. Such motility fields are generally known to result in inhomogeneous nonequilibrium distributions of active particles. This is because active particles tend to populate regions of low motility strength where they achieve longer residence times \cite{Brendel2015, Lowen2018, Cate2016, Naji2020}. We show that a more diverse picture emerges across the parameter space, and active particles can accumulate more strongly inside and outside the inclusion (depending on the motility strengths and hardness of the inclusion membrane), causing a similarly diverse range of behaviors (including sign change) for the pressure on the inclusion.
	
	The paper is organized as follows: We introduce our model in Sec. \ref{sec:model} and discuss our model predictions for distribution of active particles in Sec. \ref{sec:den} and followed by an analysis of effective pressure on the inclusion in Sec. \ref{sec:pressure}. The paper is summarized in Sec. \ref{sec:summary}.
	
	\section{Model and Methods}
	\label{sec:model}
	
\begin{figure*}[t!]
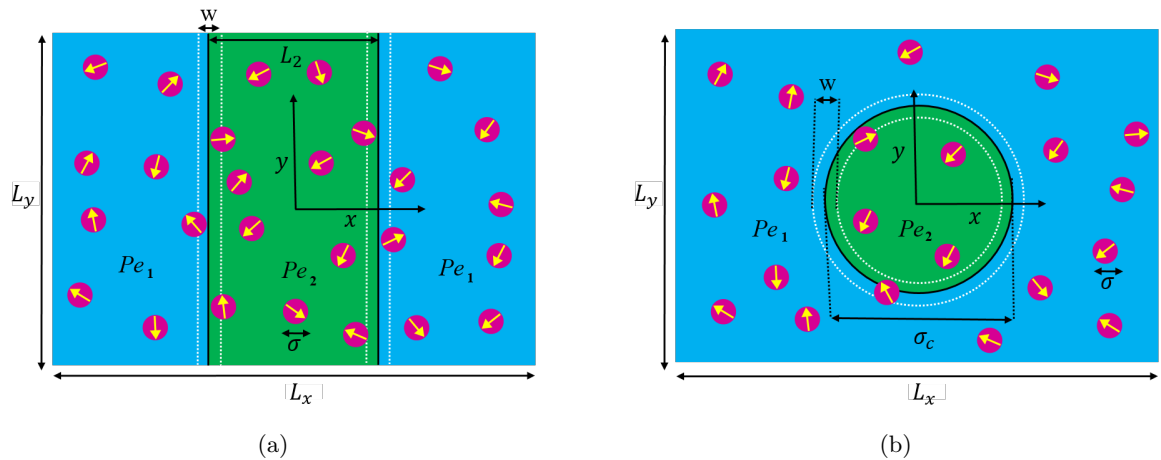

		\centering
		\begin{minipage}[b]{0.4\textwidth}
			\centering
			\includegraphics[width=\textwidth]{1_a.pdf}  \vskip 2mm (a)
		\end{minipage}
		\hskip 1cm
		\begin{minipage}[b]{0.4\textwidth}  
			\centering 
			\includegraphics[width=\textwidth]{1_b.pdf} \vskip 2mm (b)
		\end{minipage}
		\caption{Schematic view of (a) rectangular and (b) disklike permeable inclusions in a bath of active Brownian particles (of diameter $\sigma$) with mismatching P\'eclet numbers  outside ($Pe_1$) and inside ($Pe_2$) the inclusions. Dashed white lines in (a) and  dashed white circles in (b) show the inner and outer borders of permeable  membranes (of thickness $w$) that enclose each inclusion. Correspondingly, the solid black lines and circle show the midlines of the membrane.}
		\label{fig:schematic}
	\end{figure*}

	Our model consists of disklike active Brownian particles of diameter $\sigma$ moving with a constant self-propulsion speed on a 2D confining surface of area $L_x \times L_y$ \cite{Sega2019}. Realistic examples of this type of model include shaken granular matter \cite{Ramaswamy2014}, swarming bacteria \cite{Ariel2019}, and microswimmers confined to the interface of two different fluids \cite{Bishop2017}. The confining surface embeds a hollow permeable inclusion of rectangular or disk shape fixed at its center. The inclusion comprises a membrane enclosure (wall) of thickness $w$ that separates the inside and outside media, which themselves impart different activities on the particles. In the case of a rectangular inclusion, the length of the inclusion in the $y$ direction matches the entire length of the confining surface, and the centerlines of its walls (black vertical lines) are separated from each other by a distance of $L_2$ in the $x$ direction; see Fig. \ref{fig:schematic}a. In the case of a disklike inclusion, the effective diameter of the inclusion (distance between the center and middle of its membrane, shown by a black circle) is denoted by $\sigma_c$; see Fig. \ref{fig:schematic}b. The membrane enclosure is represented by a soft, repulsive potential of range $\sigma'$ (to be specified later) and thus acts as a permeable interfacial region of thickness
\begin{equation}
	w=\sigma
	\label{w}
\end{equation}
The active Brownian particles move with different self-propulsion speeds, $v_1$ and $v_2$, outside and inside the inclusion, respectively. This discontinuous motility field \cite{Naji2020, Brendel2015} is formally expressed as:
	\begin{eqnarray}
	v_\text{sp}({\mathbf r})&=&v_2+(v_1-v_2)\Theta\!\left(r'\right),\\ \nonumber\\
	r'&=&\!\left\lbrace 
	\begin{array}{ll}
	|x|-L_2/2&: \text{rectangular inclusion},\\ \\
	|{\mathbf r}|-\sigma_{c}/2&: \text{disklike inclusion},
	\end{array}
	\right.
	\label{distance}
	\end{eqnarray}
	Here, ${\mathbf r}=(x,y)$ represents the spatial coordinates with respect to the origin at the center of the confining surface, and $\Theta(\cdot)$ is the Heaviside step function, which is standardly defined as $\Theta\!\left(r'>0\right)=1$ if $r'>0$ and $0$ otherwise.
	
	The overdamped Brownian dynamics of  active particles is described  by  the Langevin equations, 
	\begin{eqnarray}
	\label{Alang1}
	&&\dot{\mathbf r}_i=v_\text{sp}({\mathbf r}_i)\,\mathbf{n}_i-\mu_{T}\frac{\partial U}{\partial{\mathbf r}_i}+\mathbf{\eta}_i(t),\\
	\label{Alang3}
	&&\dot{\theta_i}=\zeta_i(t), 
	\end{eqnarray}
	where   $\{{\mathbf r}_i(t)\}=\{(x_i(t), y_i(t))\}$ are  the position vectors and $\{{\mathbf{n}_i (t)\}=\{(\cos\theta_i(t),\sin\theta_i(t))}\}$ are the self-propulsion orientation vectors of active particles  labeled by $i=1\ldots,N$. The angular orientation $\theta_i$ is measured from the $x$ axis,  $\mu_T$ is the translational particle mobility, $U=U(\{{\mathbf r}_j\},\mathbf{R}_I)$  is the sum of the interaction potentials between the constituent particles, where $\mathbf{R}_I=(0,0)$ is the position vector of the inclusion. $\mathbf{\eta}_i(t)$ and $\mathbf{\zeta}_i(t)$ in Eqs. (\ref{Alang1}) and (\ref{Alang3})  are  translational and rotational Gaussian 
		white noises, respectively, that are characterized by zero mean, $\langle \mathbf{\eta}_i(t) \rangle=\langle \mathbf{\zeta}_i(t) \rangle=0$, and the correlators
	\begin{eqnarray}
	\label{cor1}
	&&\langle \mathbf{\eta}_i(t) \mathbf{\eta}_j(t')\rangle=2D_{T}\delta_{ij}\delta(t'-t)\\
	\label{cor2}
	&&\langle\mathbf{\zeta}_i(t)\mathbf{\zeta}_j(t')\rangle=2D_{R}\delta_{ij}\delta(t'-t), 
	\end{eqnarray}
	where $D_T$ and $D_R$ are the translational and rotational diffusion coefficients, respectively. The Einstein-Smoluchowski-Sutherland relation implies $D_{T}=\mu_{T}k_{\mathrm{B}}T$, and the low-Reynolds-number (Stokes) hydrodynamics for no-slip spheres gives  $D_{R}={3D_{T}}/{\sigma^{2}}$ \cite{Brenner1983}. 
	
	The active particles interact with each other via a short-ranged steric pair potential, utilized through the Weeks-Chandler-Andersen (WCA) pair potential, $U_{\textrm{WCA}}$; i.e., for the $i$th and $j$th active particles, we have
	\begin{eqnarray}
	\!\!\!U_{\textrm{WCA}}^{(ij)}\!=\!\left\lbrace 
	\begin{array}{ll}
	\!\epsilon\bigg[\!\!\left(\frac{\sigma}{r_{ij}}\right)^{12}\!\!-2\!\left(\frac{\sigma}{r_{ij}}\right)^{6}\!\!+\!1\bigg]
	\,\,&: \,\,r_{ij}\leq \sigma,\\ \\
	\!0 \,\,&: \,\, r_{ij}> \sigma,
	\end{array}
	\right.
	\label{WCA}
	\end{eqnarray}
	where $r_{ij}=|{\mathbf r}_i-{\mathbf r}_j|$. When active particles go through the interfacial region of the inclusion, they experience a soft repulsive WCA (sWCA) potential, which for the $i$th active particle, is assumed to be of the form \cite{Naji2020}
	\begin{equation}
	U_{\textrm{sWCA}}^{(i)} =  
	\tilde{F}_{\mathrm{max}}\,\beta\left[\frac{\sigma'^{\,12}}{({r'}_{i}^2+\alpha^{2})^6} - \frac{\sigma'^{\,6}}{({r'}_{i}^2+\alpha^{2})^3}+\frac{1}{4}\right], 
	\label{SWCA}
	\end{equation}
for $|r'_{i}|\leq\sigma'$, and $U_{\textrm{sWCA}}^{(i)} = 0$ otherwise. 	
	Here, $r'_{i}$ is defined through Eq. (\eqref{distance}) and we have
	\begin{equation}
	\begin{array}{l}
	 \sigma'=(\sigma+w)/2, \ \ \ \ \alpha=(2^{1/3}-1)^{1/2}\sigma',\\ \\
	 \beta=8.038\times10^{-5}k_\text{B}T
	 \label{beta}
	\end{array}
	\end{equation}
	where $\beta$ is a coefficient used to normalize the maximum value of the derivative of the expression enclosed by brackets in Eq. ($\eqref{SWCA}$) with respect to $r'_i$. The dimensionless parameter $\tilde{F}_{\mathrm{max}}$ represents the magnitude of the maximum force produced by the sWCA potential. In appendix \ref{app}, we demonstrate that $\tilde{F}_{\mathrm{max}}=2Pe_p$, where $Pe_p$ can be viewed as the characteristic P\'eclet number the active particles would require to successfully transit  through the inclusion membrane in the absence of thermal noise and interactions between active particles.  We utilize $Pe_p$ to identify the hardness of the inclusion membrane.

	\subsection{Simulation methods and parameters}
	
	We employ standard Brownian dynamics methods \cite{McCammon1978} to numerically solve the Eqs. (\ref{Alang1}) and (\ref{Alang3}). We calculate the effective pressure imparted on the inclusions in the steady state. The results are analyzed in terms of dimensionless P\'eclet numbers (which we may interchangeably refer to as  motility strengths) with the definitions
	\begin{equation}
	\label{Pe}
	Pe_{1,2}=\frac{\sigma\,v_{1,2}}{2D_{T}}
	\end{equation}
	for the regions outside (1) and inside (2) the inclusion, respectively, P\'eclet numbers are varied over a wide range of values from 20 up to 100. These values correspond to persistence lengths for active particles, ranging from $l_p=v/D_R=13.3\sigma$ up to $66.6\sigma$.
	
	In the simulations, we consider a rectangular simulation box with periodic boundary conditions that imitate linear and two-dimensional arrays of inclusions for the cases of rectangular and disklike inclusions, respectively. 
		For the most part, we fix the global average density of active particles as $\bar{\rho}=N\sigma^2/A=0.128$, corresponding to an area fraction of $\phi=N\pi\sigma^2/(4A)=0.1$. The surface area of the 2D simulation box is taken as $A=L_xL_y=800\sigma^2$. In the case of rectangular and disklike inclusions, we use $L_x=2L_y$ and $L_x=L_y$, respectively. We shall consider a higher global average density of active particles, $\bar{\rho}=0.255$, corresponding to $\phi=0.2$, in Sec. \ref{Sec: area fraction}, where we focus on the case of interacting active particles and examine the role of their area fraction. As defined, $\bar{\rho}$ gives the rescaled bulk density of active particles without excluding the inclusion area. The width of the rectangular inclusion is taken as $L_2=5\sigma$, $10\sigma$, and $20\sigma$, and the diameter of the disklike inclusion is denoted as $\sigma_{c}=5\sigma$, $10\sigma$, and $20\sigma$. We use a fixed value of $\epsilon=10k_{\mathrm{B}}T$ and vary $Pe_p$ from 26.3 to 78.9 to study the effects of membrane hardness.
	
	The simulations involve 100 to 204 active particles and a rescaled simulation time step of $\delta \bar{t}=D_T\delta t/\sigma^2=1.33\times 10^{-4}$. We use a total of $1.5\times 10^{7}$ to $2.5\times 10^{8}$ simulation time steps, with an initial $2.5\times10^{6}$ to $1.2\times10^{8}$ number of steps used for relaxation purposes. For lower P\'eclet numbers, we use longer simulation times because the system slowly reaches the steady state and the measured quantities slowly fluctuate around their mean values. We extend the measurement time to ensure that the noise cancels out, allowing for more accurate computation of the desired quantities. The averaged quantities are calculated by further averaging them across 4 to 10 statistically independent simulations.
	

	\label{sec: results}
\section{Spatial distribution of active particles and cross-membrane permeation}
\label{sec:den}

\begin{figure*}
	\centering
	\begin{minipage}[b]{0.405\textwidth}
		\centering
		\includegraphics[width=\textwidth]{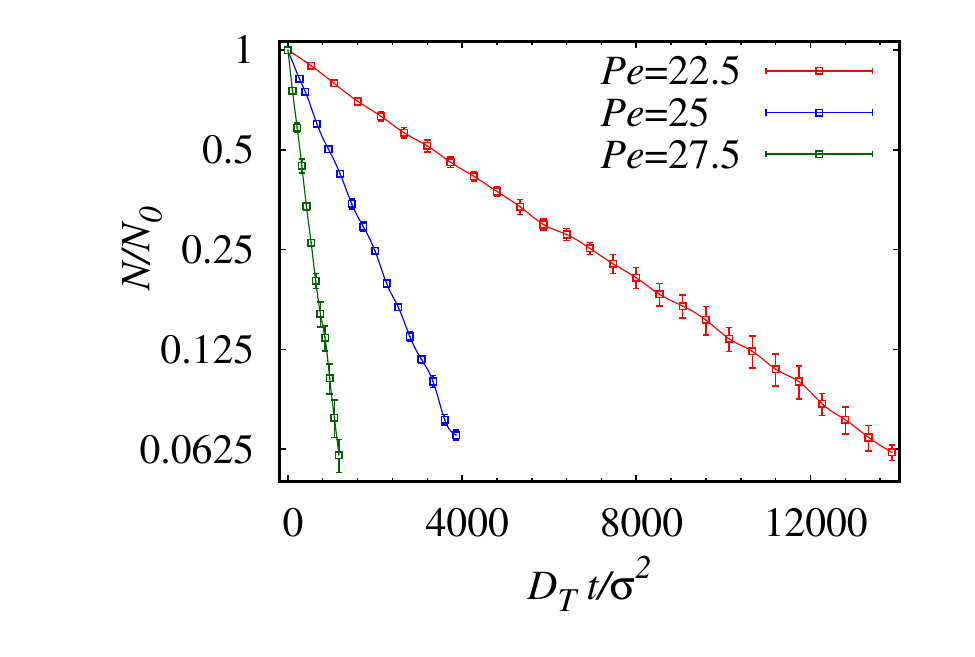}  \vskip-2mm (a)
	\end{minipage}
	\begin{minipage}[b]{0.37\textwidth}  
		\centering 
		\includegraphics[width=\textwidth]{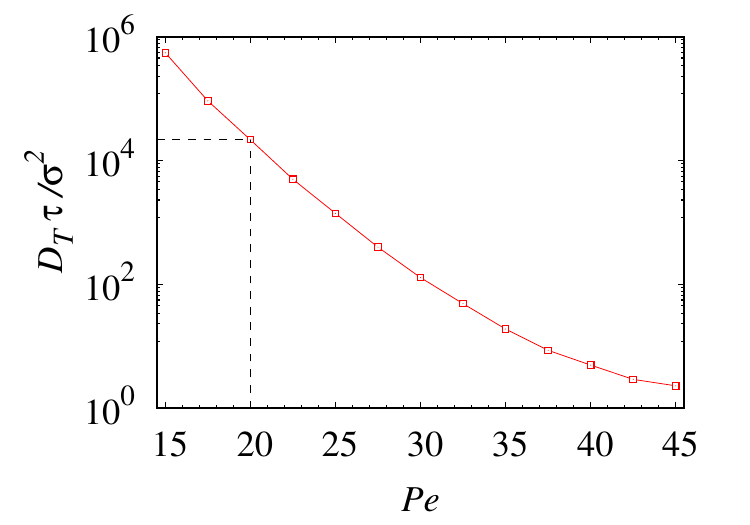} \vskip-2mm (b)
	\end{minipage}
	\caption{(a) Fraction of particles remaining inside the rectangular inclusion during depletion plotted on a logarithmic scale as a function of rescaled time for various motility strength values.
		(b) Rescaled time constant, $\tau$, see Eq. (\ref{exp_t}), plotted as a function of motility strength for the rectangular inclusion. In both panels, the width of the rectangle is $L_2=20\sigma$ and the membrane hardness is $Pe_p=25.6$. Additionally, in panel (b), the error bars are the same size as the symbols.}
	\label{dep}
\end{figure*}

In this section, we examine the penetration of active particles through the inclusion's membrane and the spatial distribution of active particles both inside and outside the inclusion. When we disregard the effects of thermal fluctuations and interactions between active particles in the Langevin equations \eqref{Alang1}, we observe that the membrane hardness strength, $Pe_p$, serves as the threshold motility strength that allows particles to cross the membrane. In this situation, the inclusion's membrane effectively resists active particles and prevents them from crossing the membrane when motility strengths are weaker than $Pe_p$. This finding aligns with previous studies \cite{Daddi2019a, Daddi2019b}, which have suggested that a threshold motility strength governs the penetration of individual active particles (without considering thermal noise in their dynamic equations) through the membrane. In contrast, in our standard minimal, active Brownian model, which accounts for thermal fluctuations and inter-particle interactions, active particles can cross the membrane even when motility strengths are below $Pe_p$. This phenomenon arises due to the facilitative effects of thermal fluctuations and particle collisions on membrane crossing.\\
\subsection{Membrane crossing}
To investigate the penetration of active particles through the enclosing membrane to the outside region, when particle interactions are disabled, we analyze a rectangular inclusion with a width of $L_2=20\sigma$ and a membrane hardness of $Pe_p=52.6$. Initially, all particles are located inside the inclusion. As particles move across the inclusion and cross the membrane, they become unable to return to the inclusion. We investigate the depletion of active particles within the inclusion and observe that the number of particles, $N$, decays exponentially over time according to the equation
\begin{equation}
	 N=N_0\exp(-t/\tau), 
	 \label{exp_t}
\end{equation}
where $N_0$ is the number of particles at $t=0$ and $\tau$ is the time constant, characterizing the time at which the number of particles drops by a factor $1/e$.

Figure \ref{dep}(a) illustrates the logarithmic fraction of particles that remain inside the inclusion over time for various motility strength values. By decreasing the strength of motility, active particles deplete the inclusion more slowly. Figure \ref{dep}(b) displays the time constant $\tau$ as a function of the motility strength $Pe$. We observe that $\tau$ increases significantly as $Pe$ decreases.

To allow for the penetration of active particles through the membrane, it is necessary for the combined effect of the active force and stochastic force to exceed the repulsive force exerted by the membrane at the particle's position. The thermal stochastic force is generated by thermal fluctuations and follows a Gaussian distribution with a mean of zero and a variance of $\sqrt{2/\delta\bar{t}}\ k_\text{B}T/\sigma=122.6k_\text{B}T/\sigma$, which is equivalent to a motility strength of $Pe_v=61.3$. Active particles are able to cross the membrane through a series of movements over multiple time steps, with the number of steps typically being less than $\tau_{R}/\delta t$ (where $\tau_{R}=1/D_R$ represents the persistence time of active particles). This is because particles must cross the membrane before their direction changes and they move away from the membrane. The probability for the combined forces (active and stochastic force) to surpass the repulsive force at the particle's position in each time step is not necessarily too small for a given motility strength. However, the probability of a successful membrane crossing event, which would be related to the product of the aforementioned single-time-step probabilities, will particularly be small. As the motility strength of particles decreases, thermal fluctuations need to generate a sequence of stronger stochastic forces in order for particles to overcome the potential barrier of the membrane. The Gaussian distribution of the thermal stochastic force makes the generation of this force sequence less likely. As a result, active particles make more frequent attempts to cross the membrane, leading to an increased membrane crossing time. Additionally, reducing the motility strength $Pe$ \cite{Gompper2013} reduces the accumulation of active particles against the membrane, resulting in a slower depletion (escape) of active particles from within the inclusion.

Figure \ref{dep}(b) indicates that the distribution of active particles inside and outside the inclusion reaches their steady state distribution more slowly from a given initial distribution as the motility strengths $Pe_1$ and $Pe_2$ decrease. To accelerate the achievement of a steady state, we homogeneously distribute the active particles throughout the simulation box. However, when it comes to extremely low motility strengths, reaching a steady state is not feasible with the computational resources currently available. Therefore, we restrict our study to intermediate and high motility strengths, where $20\leq Pe_{1,2}\leq100$, in order to analyze the system in a steady state where the computed properties have relatively constant mean values over time. The dashed lines indicate the limitations of our computational resources, where $Pe=20$ and $\tau D_T/\sigma^2=21700$. This requires more than $10^8$ time steps for the system to reach a steady state.

In the case of interacting active particles, collisions between particles further facilitate membrane crossing. Here, the time it takes for an individual particle to cross the membrane depends not only on the motility strength but also on the average density of particles inside the inclusion. One might naively expect that a constant fraction of active particles is depleted from inside the inclusion over a specific period of time, implying an exponential time dependence for the depletion process. However, due to the fact that the average speed of membrane crossing for individual particles is influenced by the time-varying interior average density of particles, we find that there is no specifically simple trend for the depletion process.

\begin{figure*}
		\centering
		\begin{minipage}[b]{0.45\textwidth}
			\centering
			\includegraphics[width=\textwidth]{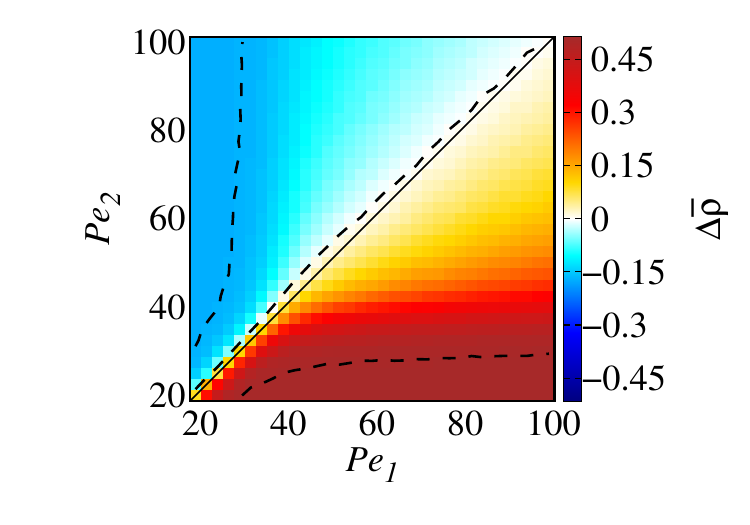}  \vskip-2mm (a)
		\end{minipage}
		\begin{minipage}[b]{0.45\textwidth}  
			\centering 
			\includegraphics[width=\textwidth]{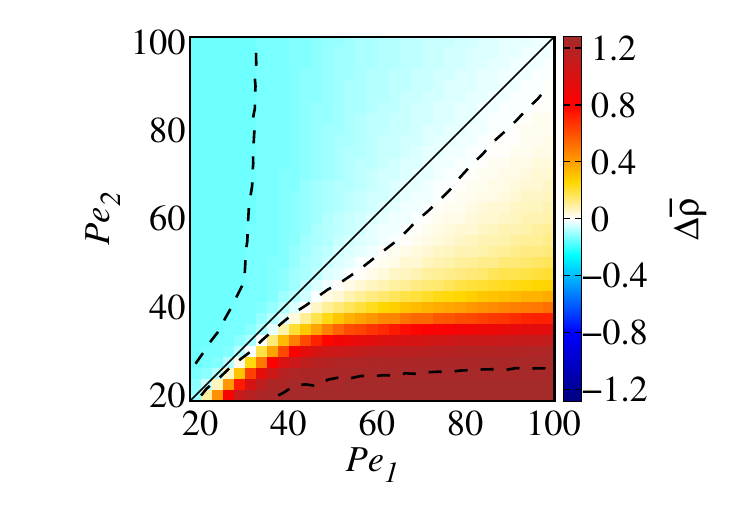} \vskip-2mm (b)
		\end{minipage}
		\vskip\baselineskip
		\begin{minipage}[b]{0.45\textwidth}   
			\centering 
			\includegraphics[width=\textwidth]{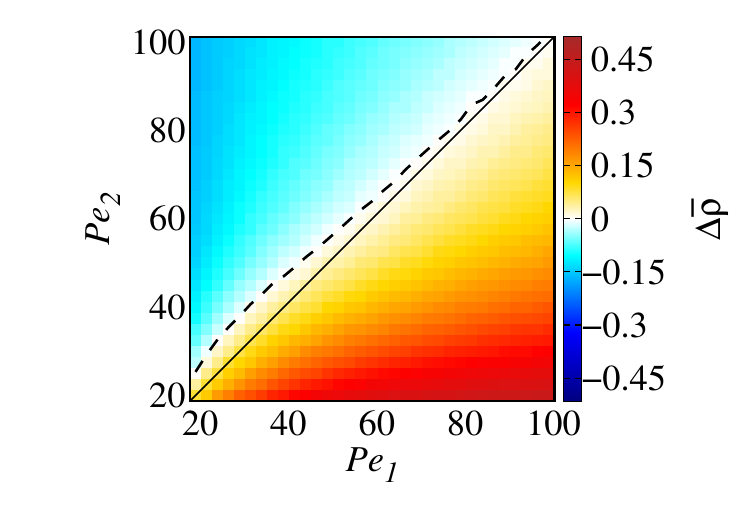} \vskip-2mm (c)
		\end{minipage}
		\begin{minipage}[b]{0.45\textwidth}   
			\centering 
			\includegraphics[width=\textwidth]{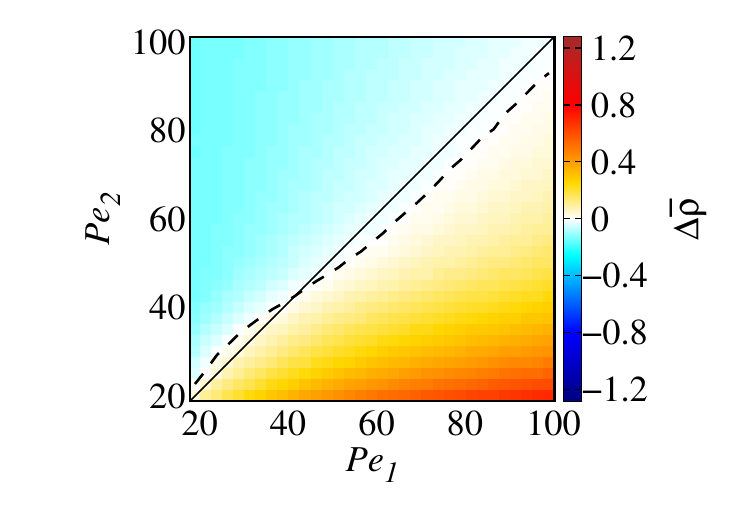} \vskip-2mm (d)
		\end{minipage}
		
		\caption{(a) The rescaled density difference between region 2 (inside the inclusion) and region 1 (outside the inclusion) is shown as a color-coded density map in terms of $Pe_1$ and $Pe_2$ for noninteracting active particles in a rectangular inclusion with a width of $L_2=10\sigma$. (b) Same as (a) but here we show the results for a disklike inclusion with a diameter of $\sigma_c=10\sigma$. Panel (c) presents the density difference for interacting active particles in a rectangular inclusion with a width of $L_2=10\sigma$. Panel (d) depicts the rescaled density differences for a disklike inclusion with a diameter of $\sigma_c=10\sigma$.  In all panels, the parameters $\bar{\rho}=0.128$ and $Pe_p=52.6$ are maintained. The middle dashed line represents the contour line of zero density difference. The dashed lines within the red and blue areas indicate the contour lines of density difference for values of $0.99\Delta\bar{\rho}_{trp}$ and $0.99\Delta\bar{\rho}_{dep}$, respectively (the latter lines are only visible in panels (a) and (b). The solid diagonal lines depict the reference bisector $Pe_1=Pe_2$.}
		\label{den}
	\end{figure*}
	
\subsection{Spatial distribution of active particles}
	
To investigate the densities of active particles within and outside the inclusion (regions 2 and 1) across different motility strengths $Pe_1$ and $Pe_2$, we calculated the difference in average density between the two regions, denoted as $\Delta\rho=\rho_2-\rho_1$. Here, $\rho_{1,2}=N_{1,2}/(L_{1,2}L_y)$, where $N_{1,2}$ represents the number of particles present in region 1 (outside the inclusion) and region 2 (inside the inclusion), and $L_{1,2}$ denotes the corresponding region widths. Figure \ref{den}(a) presents a color map illustrating the difference in rescaled average densities, $\Delta\bar{\rho}$, as a function of $Pe_1$ and $Pe_2$ for a rectangular inclusion with a width of $L_2=10\sigma$ and noninteracting particles. Contour lines of constant density difference, obtained by interpolating discrete data acquired at resolutions of $\Delta Pe_{1}=2.5$ and $\Delta Pe_{2}=2.5$, are represented by dashed lines. Positive density differences are indicated by warm colors, while cool colors signify negative density differences. As depicted in Fig. \ref{den}a, regions with lower motility generally exhibit higher average density due to particles spending more time in these areas and facing increased difficulty in crossing the membrane. Greater disparities in motility strengths result in more significant variations in density. The range of rescaled density differences, $\Delta \bar{\rho}$, spans from $-0.17$ to $0.51$, indicating complete particle depletion from the inclusion at the lower end of the range and complete particle trapping within the inclusion at the higher end. Positive density differences vary over a wider range compared to negative differences, which can be attributed to the smaller area of the inclusion leading to a higher average density. By reducing the width of the inclusion $L_2$, the range of positive density differences increases while the range of negative density differences decreases. The dashed lines in the blue and red regions of the color map shown in Fig. \ref{den}a represent the parameter space $\{Pe_1, Pe_2\}$ where particles are nearly depleted from the inclusion and almost completely trapped within the inclusion, respectively. The dashed line on the border of the red and blue areas represents the contour line of zero density difference. Notably, this line is positioned above the reference line, denoted as $Pe_1=Pe_2$, which we refer to as the reference line for brevity. The reference line is depicted as a solid black line in Fig. \ref{den}. For $L_2=L_x/2=20\sigma$, where the areas inside and outside the inclusion are equal, the zero contour line coincides with the reference line. As $L_2$ decreases from $20\sigma$ (data not displayed), the contour line gradually shifts further above the reference line.

Figure \ref{den}(b) displays the density difference for a circular inclusion with a diameter of $\sigma_c=10\sigma$ in the case of noninteracting particles. The range of density difference varies significantly, from $-0.142$ (density difference in full depletion, $\Delta\bar{\rho}_{dep}$) to $1.274$ (density difference in full trapping, $\Delta\bar{\rho}_{trp}$), which is considerably wider than the range observed for the rectangular inclusion of $L_2=10\sigma$. This is because the circular inclusion has a smaller area than the rectangular inclusion. As a result, when active particles are fully trapped within the inclusion, there is a greater density difference. In contrast to the rectangular inclusion, the contour line representing zero density difference lies below the reference line. This characteristic becomes increasingly prominent for higher motility strengths.

Figures \ref{den} (c) and \ref{den}(d) display the density differences for interacting particles in the case of a rectangular inclusion with a width of $L_2=10\sigma$ and a circular inclusion with a diameter of $\sigma_c=10\sigma$, respectively. As shown in the panels, the density differences exhibit weaker variations within a narrower range (ranging from $-0.162$ to $0.437$ for the rectangular inclusion and from -0.140 to 0.691 for the disklike inclusion) compared to the density differences observed in the noninteracting case depicted in Figs. \ref{den}(a) and \ref{den}(b). This is because collisions between interacting particles facilitate the crossing of membranes, thereby reducing the difference in residence time of active particles between regions 1 and 2. However, it is important to note that there are limitations on the average particle density inside or outside the inclusion when dealing with interacting particles. Excluded volume interactions between particles prevent the density from exceeding the density of hexagonal close packing, $\bar{\rho}_{hcp}=1.159$, which corresponds to a packing fraction of $\phi_{hcp}=0.91$ \cite{Baskaran2013}. For the disk with a diameter of $\sigma_c=10\sigma$, active particles cannot be completely trapped inside the inclusion due to the high average density of particles inside the inclusion, which exceeds $\bar{\rho}_{hcp}$. The excluded volume interaction allows for a maximum difference in density of $\Delta\bar{\rho}=1.146$. Figure \ref{den}(d) shows that the maximum density difference is not achieved for motility strengths within the range of $20\leq Pe_{1,2}\leq 100$. In Fig. \ref{den}(c), which illustrates the density difference for a rectangular inclusion with a width of $L_2=10\sigma$ and interacting particles, the zero contour line (dashed line) lies higher above the reference line ($Pe_1=Pe_2$) compared to the noninteracting case. The dashed line in Fig. \ref{den}(d) indicates the contour line for zero density difference of a disk-shaped inclusion with a diameter of $\sigma_c=10\sigma$ in the case of interacting particles. By increasing $Pe_1$, the zero contour line initially shifts above the reference line, but eventually shifts below it.

\begin{figure*}
		\centering
		\begin{minipage}[b]{0.37\textwidth}
			\centering
			\includegraphics[width=\textwidth]{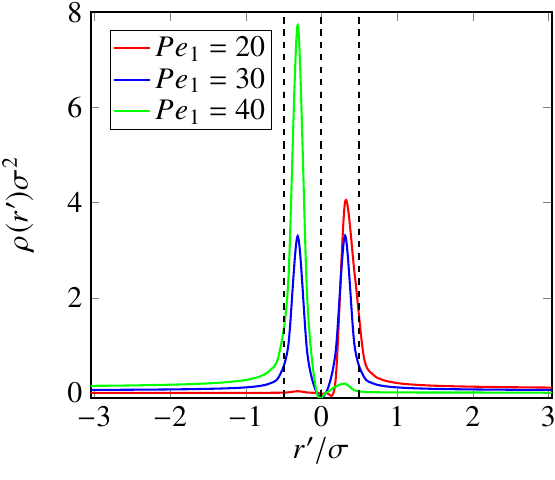}  \vskip-2mm (a)
		\end{minipage}
		\hskip1cm
		\begin{minipage}[b]{0.37\textwidth}  
			\centering 
			\includegraphics[width=\textwidth]{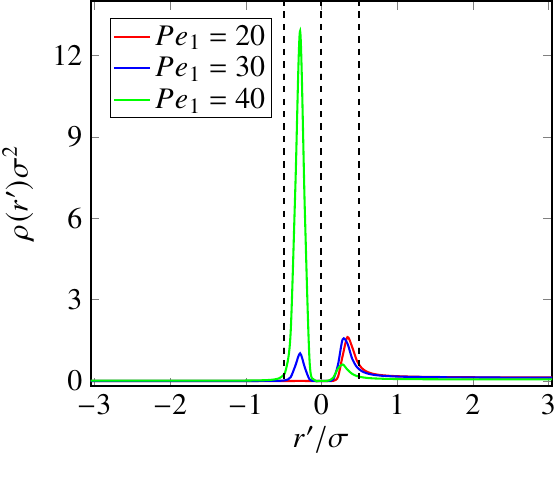} \vskip-2mm (b)
		\end{minipage}
		\vskip\baselineskip
		\begin{minipage}[b]{0.37\textwidth}   
			\centering 
			\includegraphics[width=\textwidth]{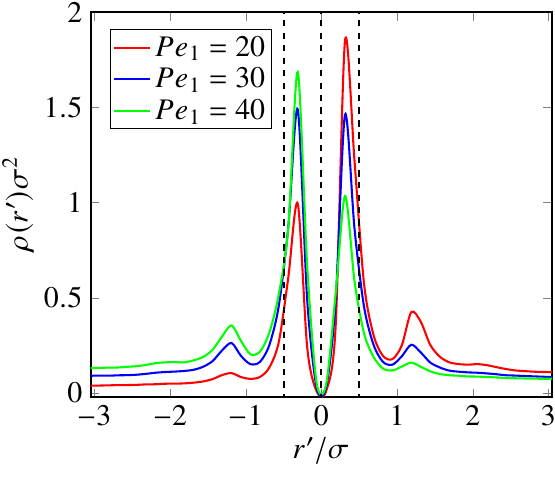} \vskip-2mm (c)
		\end{minipage}
		\hskip 1cm
		\begin{minipage}[b]{0.37\textwidth}   
			\centering 
			\includegraphics[width=\textwidth]{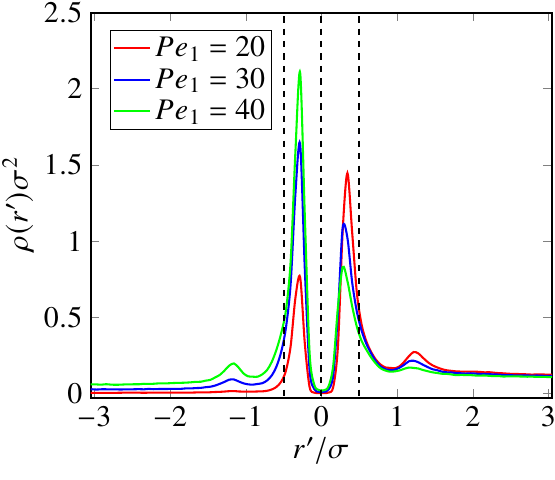} \vskip-2mm (d)
		\end{minipage}
		\caption{Rescaled density profiles of active particles as a function of rescaled distance from the inclusion membrane with a membrane hardness of $Pe_p=52.6$ are shown for fixed $Pe_2=30$ and different values of $Pe_1$ as indicated on the graphs. $r'$, distance from the inclusion membrane is defined as Eq. (\ref{distance}). For the rectangular inclusion, $r'$ is the lateral position (distance) relative to the midline of the inclusion membrane while for the disklike inclusion, the radial position (distance) from the middle circle of the inclusion membrane (see Fig. \ref{fig:schematic}). The plus and minus signs of the distance indicate whether the position is outside or inside the inclusion, respectively. Panel (a) displays the density profile for noninteracting active particles in a rectangular inclusion with a width of $L_2=10\sigma$, while panel (b) shows the density profile for noninteracting active particles in a disklike inclusion with a diameter of $\sigma_c=10\sigma$. Panel (c) illustrates the density profile for interacting active particles in a rectangular inclusion with a width of $L_2=10\sigma$, and panel (d) presents the density profile for interacting active particles in a disklike inclusion with a diameter of $\sigma_c=10\sigma$. The interfacial region around the inclusions is indicated by three dashed lines.} 
		\label{xrden}
	\end{figure*}

Fig. \ref{den} gives an overall view of how the active particles are distributed in both regions. To gain a more comprehensive understanding of their distribution, we analyze the density of active particles near the inclusion membrane and in the bulk, both in the interior and exterior regions. Since, in the distribution of active particles, there is axial symmetry relative to the $y$ axis in the case of rectangular inclusion and circular symmetry in the case of disklike inclusion, we examine the density profile based on the distance from the midline of the inclusion membrane which we denote by $r'$ [see Eq. (\ref{distance})]. The value of $r'$ represents the nearest distance from the membrane, with the plus and minus signs referring to positions located outside ($+$) or inside ($-$) the inclusion. Figures \ref{xrden}(a) and \ref{xrden}(b) display the density profile of noninteracting active particles for a rectangular inclusion with a width of $L_2=10\sigma$ and a disk-shaped inclusion with a diameter of $\sigma_c=10\sigma$, respectively. Figures \ref{xrden}(c) and \ref{xrden}(d) display the density profiles for interacting particles in the case of a rectangular inclusion with a width of $L_2=10\sigma$ and a disk-shaped inclusion with a diameter of $\sigma_c=10\sigma$, respectively. We set the motility strength inside the inclusion to $Pe_2=30$ and adjust the motility strength outside, $Pe_1$, within the range of 20 to 40. This range corresponds to a persistence length, $l_p$, that is 1.33 to 2.66 times larger than the width or diameter of the inclusion. In cases where the persistence lengths exceed the size of the inclusion, active particles tend to accumulate strongly against the membrane. This is reflected by significant peaks in the density profiles at distances near the membrane, as shown in all panels of Fig. \ref{xrden}. By comparing Figs. \ref{xrden}(c) and \ref{xrden}(d) (interacting case)
with Figs. \ref{xrden}(a) and \ref{xrden}(b) (noninteracting case), we observe that excluded volume interactions between the particles result in smaller peaks near the membrane. These interactions also induce a second peak in the density profile due to the layering of active particles around the membrane \cite{Naji2020,Naji2017,Naji2021c,Bolhuis2015}. In both regions (inside and outside the inclusion), the density profile displays one or more peaks that decrease as the distance from the membrane increases, eventually reaching a constant value in the bulk. The overall trend observed in all panels is that the density near the membrane and in the bulk is higher in the region of lower motility strength compared to the region of higher motility strength.

Figures \ref{xrden}(a) and \ref{xrden}(c) depict the density profile of  noninteracting and interacting active particles, respectively, in the case of a rectangular inclusion with a width of $L_2=10\sigma$. The blue curves in Figs. \ref{xrden}(a) and \ref{xrden}(c) indicate that the densities are almost the same near the membrane and in the bulk for $Pe_1=Pe_2=30$. However, Figs. \ref{den}(a) and \ref{den}(c) show that the density difference between the two regions is positive, indicating that the average density of active particles inside the rectangular inclusion is higher than outside. This discrepancy can be explained as follows: the average densities inside and outside the inclusion are equal to $2\int_{-L_2/2}^{0}(\rho(r')-\rho^{bulk}_{2})dr'/L_2+\rho^{bulk}_{2}$ and $2\int_{0}^{L_1/2}(\rho(r')-\rho^{bulk}_{1})dr'/L_1+\rho^{bulk}_{1}$, respectively. Here, $\rho^{bulk}_{1,2}$ represent the density of active particles in the bulk (i.e., sufficiently away from the boundaries) for regions 1 and 2 (i.e., outside and inside the inclusion, respectively). These bulk densities are almost the same in both regions.  However, $[\rho(r')-\rho^{bulk}_{1,2}]$ equals zero everywhere except near the membrane in both regions. As a consequence, the foregoing spatial integrals are equal to the number of active particles accumulated near the membrane per unit of the membrane length ($L_y$) \cite{Gompper2013}. Also, the number of particles accumulated near the membrane, both inside and outside the inclusion, are equal. This is because the density peaks near the membrane have the same heights and widths in both regions. Since $L_2<L_1$, the average density inside the inclusion is larger than outside. For large motility strengths, the accumulation of active particles against the membrane is weakened due to likelier membrane crossing events. This is the reason why we see a decrease in the density difference around the reference line $Pe_1=Pe_2$ (shown by the  solid black line) in Figs. \ref{den}(a) and \ref{den}(c). Additionally, we observe that the contour line of zero density difference gets closer to the reference line at large motility strengths.

Figures \ref{xrden}(b) and \ref{xrden}(d) display the density profile for a disklike inclusion with a diameter of $\sigma_c=10\sigma$ in the case of noninteracting and interacting active particles, respectively. For $Pe_1=Pe_2=30$, inside the inclusion, the densities of particles near the membrane and in the bulk are lower than outside for the case of noninteracting particles [see blue curve in Fig. \ref{xrden}(b)]. For the case of interacting particles, these are reversed [see blue curve in Fig. \ref{xrden}(d)].
\\
\section{Effective Pressure on the Inclusion}
\label{sec:pressure}
We thoroughly investigate the roles of system parameters on the effective pressure acting on the inclusion. These parameters include the mismatch in motility fields between the inside and outside regions of the inclusion, the geometry of the inclusion, the excluded volume interaction between active particles, the membrane hardness, as well as the global average density of active particles on the effective pressure. The effective pressure on the inclusion membrane is obtained following its mechanical definition \cite{Kardar2015-2} and is expressed in rescaled form as:

\begin{equation}
	\bar{P}=\frac{1}{\rho lk_\text{B}T}\sum_{i=1}^{N}\left(2\Theta\!\left(r'_i\right)-1\right)\bigg\langle\left|\frac{\partial}{\partial {{\mathbf r}_i}}U_{\textrm{sWCA}}^{(i)}\right|\bigg\rangle
	\label{eq:pres}
\end{equation}

where $l$ is the perimeter of the membrane enclosure, $\rho$ is the global average density of active particles, and $r'_i$ is defined by Eq. (\eqref{distance}), and the brackets denote steady-state time averaging over different simulation runs. The expression inside the brackets represents the instantaneous force due to the $i$th active particle as it interacts with the inclusion membrane. $\bar{P}>0$ ($\bar{P}<0$) indicates an outward (inward) pressure.

\subsection{noninteracting active particles}
\label{sec:NI}
\begin{figure*}[t!]\centering
		\hskip-7mm
		\\
		\begin{minipage}[t]{0.36\textwidth}\begin{center}
				\includegraphics[width=\textwidth]{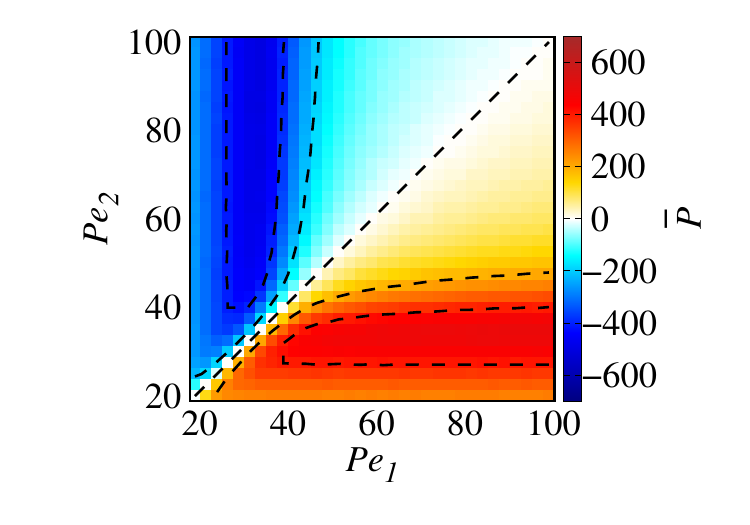} \vskip-2mm (a)
		\end{center}\end{minipage}\hskip-0.8cm
		\begin{minipage}[t]{0.36\textwidth}\begin{center}
				\includegraphics[width=\textwidth]{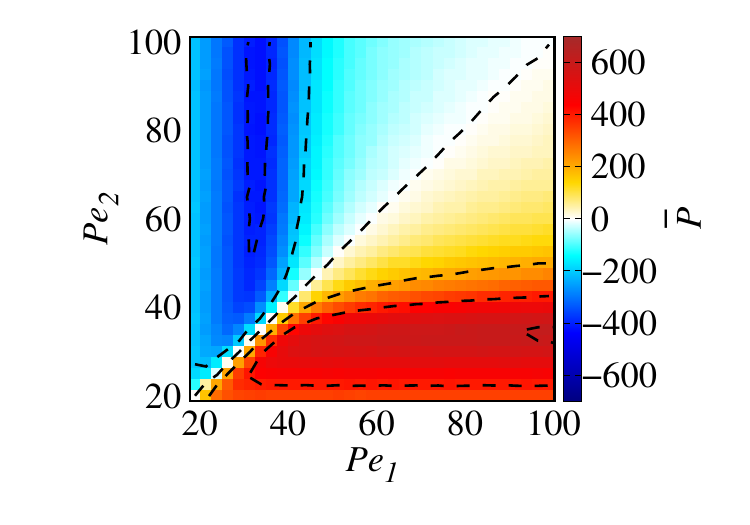} \vskip-2mm (b)
		\end{center}\end{minipage}\hskip-0.8cm
		\begin{minipage}[t]{0.36\textwidth}\begin{center}
				\includegraphics[width=\textwidth]{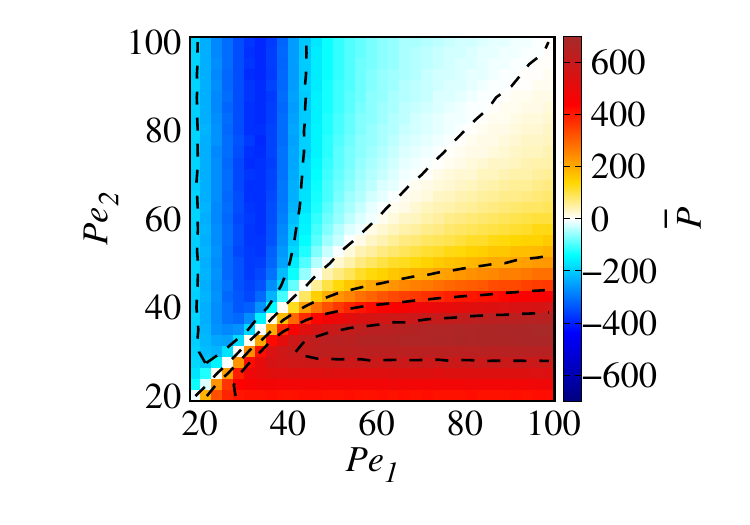} \vskip-2mm (c) 
		\end{center}\end{minipage}
		\\
		\begin{minipage}[t]{0.36\textwidth}\begin{center}
				\includegraphics[width=\textwidth]{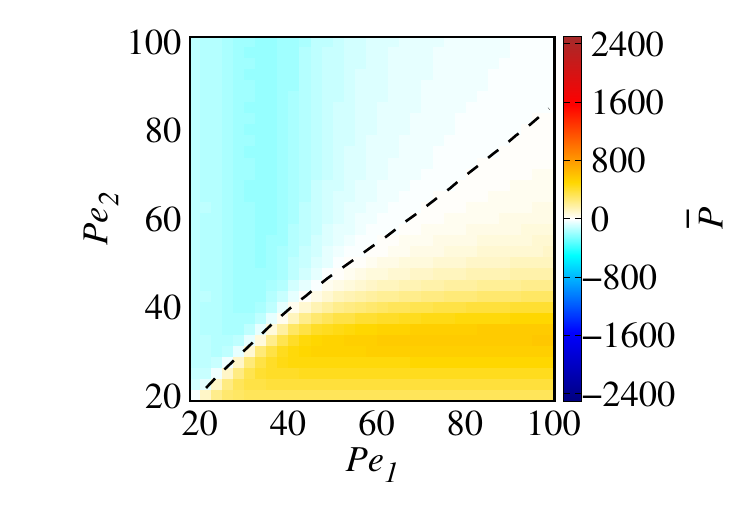} \vskip-2mm (d)
		\end{center}\end{minipage}\hskip-0.8cm
		\begin{minipage}[t]{0.36\textwidth}\begin{center}
				\includegraphics[width=\textwidth]{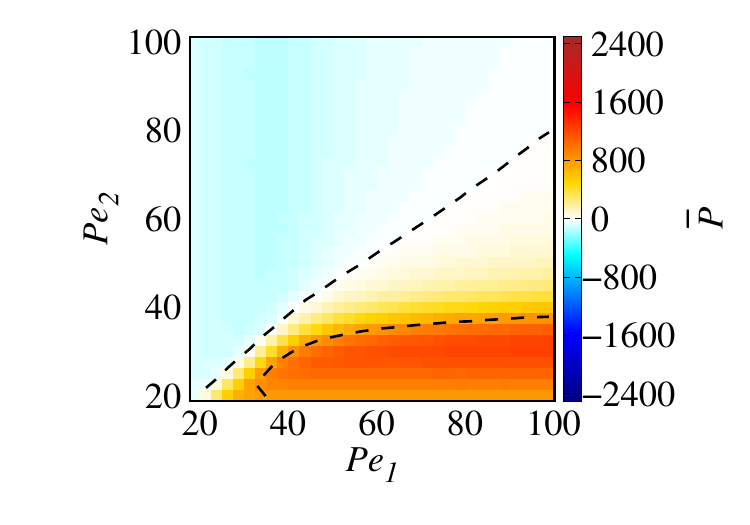} \vskip-2mm (e)
		\end{center}\end{minipage}\hskip-0.8cm
		\begin{minipage}[t]{0.36\textwidth}\begin{center}
				\includegraphics[width=\textwidth]{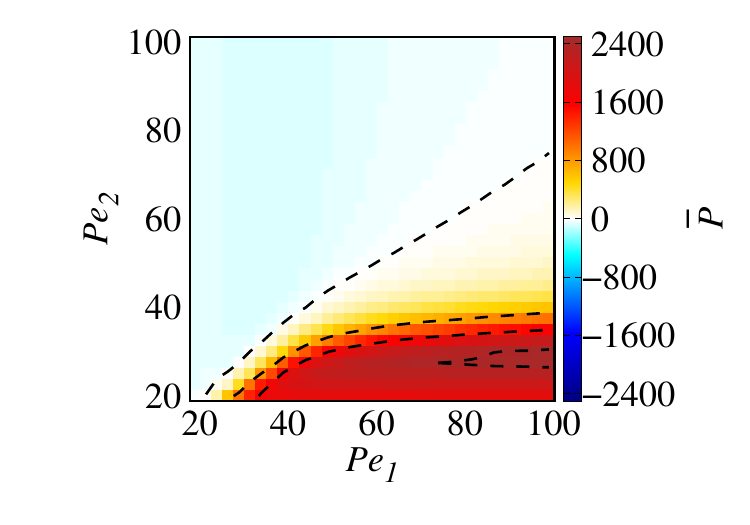} \vskip-2mm (f) 
		\end{center}\end{minipage}
		\caption{Rescaled effective pressure acting on a model inclusion as a function of Péclet numbers $\{Pe_1, Pe_2\}$ for noninteracting active particles and enclosing membrane hardness $Pe_p=52.6$. The first row of panels displays the effective pressure on a  rectangular inclusion with different rescaled widths: (a) $L_2/\sigma=20$, (b) $10$, and (c) $5$. The second row of panels shows the effective pressure on a disklike inclusion with different rescaled diameters: (d) $\sigma_c/\sigma=20$, (e) $10$, and (f) $5$. Dashed lines represent pressure contours at the tick values of the color sidebar for each panel.}
		\label{NRec}
	\end{figure*}

\begin{figure*}
		\centering
		\begin{minipage}[b]{0.37\textwidth}
			\centering
			\includegraphics[width=\textwidth]{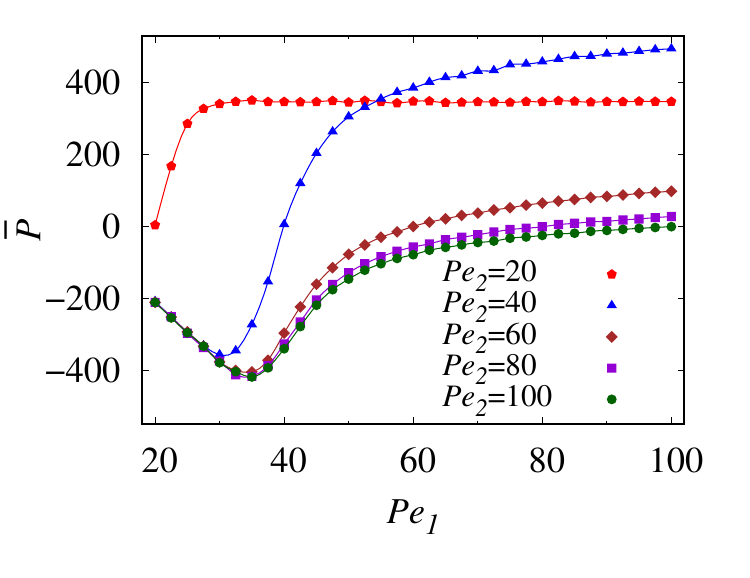}  \vskip-2mm (a)
		\end{minipage}
		\begin{minipage}[b]{0.37\textwidth}  
			\centering 
			\includegraphics[width=\textwidth]{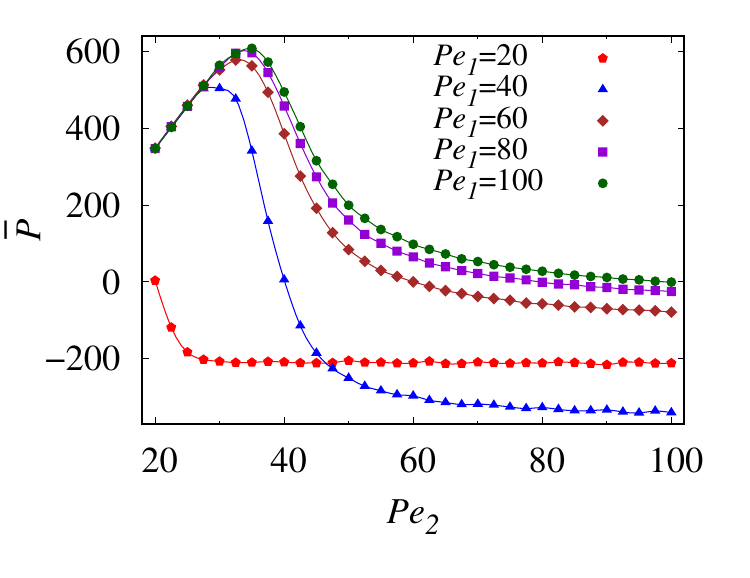} \vskip-2mm (b)
		\end{minipage}
		\caption{(a) Rescaled effective pressure acting on the rectangular inclusion, $\bar{P}$, as a function of $Pe_1$ for different fixed values of $Pe_2$, and (b) as a function of $Pe_1$ for different fixed values of $Pe_2$, as shown in the graphs. In both panels, $L_2/\sigma=10$ and enclosing membrane hardness, $Pe_p=52.6$. Errors are smaller than the size of the symbols.}
		\label{cPe}
\end{figure*}

\begin{figure}[t!]
		\begin{center}
			\includegraphics[width=1\textwidth]{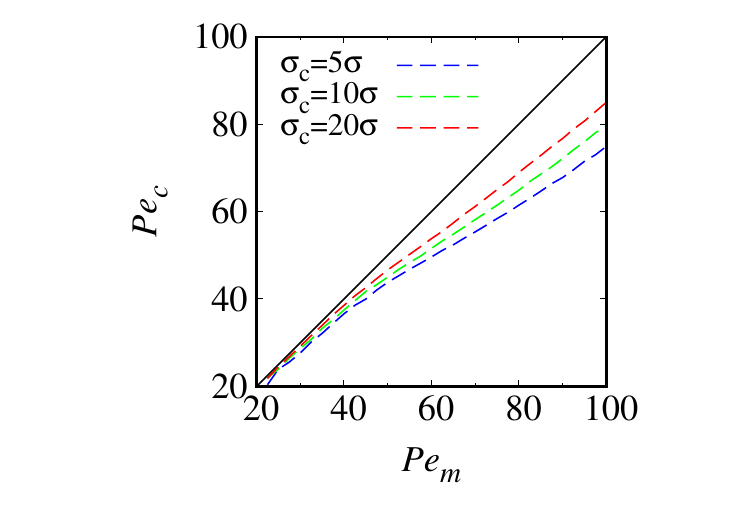}
			\caption{Contour lines of $\bar{P}=0$ as a function of $Pe_1$ and $Pe_2$ for disklike inclusions of different diameters, as indicated on plot. The solid line represents the reference line $Pe_2=Pe_1$.}
			\label{ncl0}
		\end{center}
	\end{figure}

	 Figures \ref{NRec}(a)-\ref{NRec}(c) presents the effective pressure exerted on a rectangular inclusion as a function of motility strengths, $Pe_1$ and $Pe_2$, ranging from $20$ to $100$ for noninteracting active particles with different widths of the inclusion, $L_2$, in the range of $[5\sigma, 20\sigma]$. This range of motility strengths corresponds to persistence lengths that vary from $13.3\sigma$ to $66.7\sigma$. The color scheme in the figure uses red to indicate positive (outward) pressures and blue for negative (inward) pressures. Dashed lines represent constant-pressure contour lines calculated by interpolating discrete data obtained with resolutions of $\Delta Pe_{1}=2.5$ and $\Delta Pe_{2}=2.5$.

	The effective pressure on a rectangular inclusion is primarily determined by the difference in average particle densities between the two regions and the strengths of motility fields. As discussed in the previous section, the density difference between region 2 (inside the inclusion) and region 1 (outside the inclusion) is positive when the motility strength inside the inclusion is less than that outside ($Pe_2<Pe_1$), owing to the stronger concentrate of active particles in the region of lower motility. This density difference plays a dominant role in determining the sign of the effective pressure, which is positive when $Pe_2<Pe_1$ and negative in the reversed situation ($Pe_2>Pe_1$). For $Pe_1=Pe_2$, the density difference is positive when $L_2<L_x/2=20\sigma$ [as shown in Fig. \ref{den}a], but the density of active particles near the inclusion membrane is almost equal in both regions [as depicted in the blue curve in Fig. \ref{xrden}(a)]. This leads to zero effective pressure on the rectangular inclusion [as observed in Figs. \ref{NRec}(a)-\ref{NRec}(c)].
	 
	 Figure \ref{cPe}(a) illustrates the rescaled effective pressure on a rectangular inclusion with width $L_2=10\sigma$ in relation to exterior motility strength, $Pe_1$, for various fixed interior motility strengths, $Pe_2$. When $Pe_2=20$, the effective pressure is consistently positive due to $Pe_2\leq Pe_1$ within the given range of $Pe_1$, resulting in a positive density difference. As $Pe_1$ increases, the effective pressure monotonically increases until it reaches a constant value. This is because an increase in $Pe_1$ leads to a rise in density difference until the maximum is reached, trapping total active particles within the inclusion. At this point, the exterior active pressure vanishes while the interior active pressure remains constant due to the fixed interior motility strength. Conversely, when $Pe_2>20$, the effective pressure initially decreases as $Pe_1$ increases until it reaches a global minimum at $Pe_1^*$, after which it increases. Figure \ref{cPe}(a) indicates that by increasing $Pe_2$, the global minimum shifts to the right and $Pe_1^*$ increases up to 35.
	 
	 Although the density difference reaches its minimum value at the lowest level of exterior motility strength, $Pe_1=20$, the effective pressure has a minimum value at $Pe_1^*>20$. This is because, for $Pe_1\leq Pe_1^*$, the density difference remains almost constant with $Pe_1$, and its value is close to the minimum where active particles are completely depleted from the inclusion. In this situation, the exterior active pressure increases with $Pe_1$ due to stronger interaction between exterior active particles and the inclusion membrane, resulting in a decrease in effective pressure. However, for $Pe_1>Pe_1^*$, the density difference increases (the density outside the inclusion decreases) as $Pe_1$ increases, and this effect is stronger than the effect of increasing the strength of interaction between exterior active particles and the membrane. As a result, there is an increase in effective pressure.
	 
	 The relationship between effective pressure and interior motility strength, while holding exterior motility strength $Pe_1$ constant, exhibits an opposite trend compared to the relationship between effective pressure and exterior motility strength, $Pe_1$ at fixed interior motility strengths {compare Figs. \ref{cPe}(b) and \ref{cPe}(a)}. This is due to the fact that as $Pe_2$ increases, the density difference decreases, whereas it increases with $Pe_1$. Figure \ref{cPe}(b) shows the global maxima for various fixed exterior motility strengths, while Fig. \ref{cPe}(a) displays the global minima for different fixed interior motility strengths. The maxima have a larger absolute value than the minima because the inclusion area is smaller than the surrounding region, leading to a higher average density of active particles inside the inclusion.      
	 
	  Returning to Fig. \ref{NRec}(a), we present the effective pressure for a rectangular inclusion width of $L_2=L_x/2=20\sigma$. The red and blue colors and contour lines exhibit symmetry with respect to the line $Pe_1=Pe_2$, indicating that exchanging $Pe_1$ and $Pe_2$ reverses the direction of the effective pressure. This is due to the fact that the regions inside and outside the inclusion have equal widths of $L_2=L_1=L_x/2$. However, this symmetry is disrupted for $L_2\neq L_x/2$, as shown in Figs. \ref{NRec}(b) and \ref{NRec}(c). By decreasing the inclusion width, $L_2$, outward (positive) and inward (negative) pressures become stronger and weaker, respectively, as shown from Figs. \ref{NRec}(a) to \ref{NRec}(c); the red colors become darker while blue colors become lighter. Tables \ref{max} and \ref{min} provide the maximum and minimum pressure values and their corresponding coordinates in the parameter space $\{Pe_1,Pe_2\}$ for Figs. \ref{NRec}(a)-\ref{NRec}(c).
	 
	 \begin{table}[ht]
	 	\centering 
	 	\begin{tabular}{ | l | l | l |l |}
	 		\hline
	 		Panels&$\bar{P}_{\mathrm{max}}$&$Pe_1$&$Pe_2$\\ \hline
	 		(a) $L_2/L_x=0.5$&495&100&35\\ \hline
	 		(b) $L_2/L_x=0.25$&608&100&35\\ \hline
	 		(c) $L_2/L_x=0.125$&687&100&35\\ \hline
	 	\end{tabular}
	 	\caption{Maximum pressure $\bar{P}_{\mathrm{max}}$, and its coordinates in the parameter space  $\{Pe_1, Pe_2\}$ for panels (a-c) Fig. \ref{NRec}.}
	 	\label{max}
	 \end{table}
	 
	 \begin{table}[ht]
	 	\centering 
	 	\begin{tabular}{ | l | l | l |l |}
	 		\hline
	 		Panels&$\bar{P}_{\mathrm{min}}$&$Pe_1$&$Pe_2$\\ \hline
	 		(a) $L_2/L_x=0.5$&-496&35&100\\ \hline
	 		(b) $L_2/L_x=0.25$&-419&35&100\\ \hline
	 		(c) $L_2/L_x=0.125$&-389&35&100\\ \hline
	 	\end{tabular}
	 	\caption{Minimum pressure $\bar{P}_{\mathrm{min}}$ and its coordinates in the parameter space $\{Pe_1, Pe_2\}$ for panels (a-c) Fig. \ref{NRec}.}
	 	\label{min}
	 \end{table}
	 
	The coordinates of the maximum and minimum pressure remain fixed in the parameter space $\{Pe_1,Pe_2\}$ for different inclusion widths, $L_2$. However, the values of the maximum and minimum pressures change with $L_2$. As $L_2$ decreases, the increase rate of the maximum pressure, $\bar{P}_{\mathrm{max}}$, exceeds the decrease rate of the minimum pressure, $\bar{P}_{\mathrm{min}}$. This rate of change in the maximum and minimum pressures depends on the rate of variation in the difference of the average densities in the two interior and exterior regions. By reducing the width of the region of lower (higher) motility, the average densities of active particles in both regions increase (decrease), and the absolute value of difference between them increases (decreases). Consequently, positive (outward) and negative (inward) pressures increase and decrease, respectively.
	 
	We now discuss the scenario of a disk-shaped inclusion present in an active bath. The effective pressure in this case is somewhat similar to that of a rectangular inclusion, but there are also some subtle differences that arise (as can be observed by comparing the bottom row of panels with the top ones in Fig. \ref{NRec}). The contour lines in the case of a disk-shaped inclusion differ from those in the rectangular case. Figure \ref{ncl0} displays the contour lines of zero pressure for disk-shaped inclusions of varying diameters, namely $\sigma_{c}/\sigma=20,10$, and $5$. Unlike the rectangular case, the contour lines of zero pressure for the disk-shaped inclusion do not align with the straight line of $Pe_1=Pe_2$ (which we refer to as the reference line and illustrate by the solid black line in Fig. \ref{ncl0}). The contour lines lie entirely below the reference line because the density difference is negative for $Pe_1=Pe_2$ in the case of noninteracting active particles and disk-shaped inclusion [as demonstrated in Fig. \ref{den}(b)]. For low and moderate motility strengths, the contour line of zero pressure almost coincides with the contour line of zero density difference. For a larger inclusion diameter, both contour lines match each other over a more extensive range of motility strengths (although this data is not shown). However, for sufficiently high motility strengths, the contour line of zero pressure lies below the contour line of zero density difference. This is because the density difference becomes zero when $Pe_1$ is moderately greater than $Pe_2$ (see the middle dashed line in Fig. \ref{den}(b), resulting in a moderately stronger interaction between active particles outside the inclusion and the membrane than inside the inclusion. As a result, the effective pressure is negative when the density difference is equal to or slightly more than zero.
	 
	 Figures \ref{NRec}(d)-\ref{NRec}(f) demonstrate that the effective pressure increases more significantly with a decrease in size (diameter) for disklike inclusions compared to rectangular inclusions [as seen in Figs. \ref{NRec}(a)-\ref{NRec}(c)]. This is due to the fact that the area of the disk inclusion varies proportionally with the square of its diameter, while the area of the rectangular inclusion varies linearly with its width. As a result, the density difference of the disklike inclusion increases more strongly as the diameter of the inclusion decreases.
 
	\subsection{Interacting active particles}
	\label{sec:rec}

	\begin{figure*}[t!]\centering
		\hskip-7mm
		\\
		\begin{minipage}[t]{0.37\textwidth}\begin{center}
				\includegraphics[width=\textwidth]{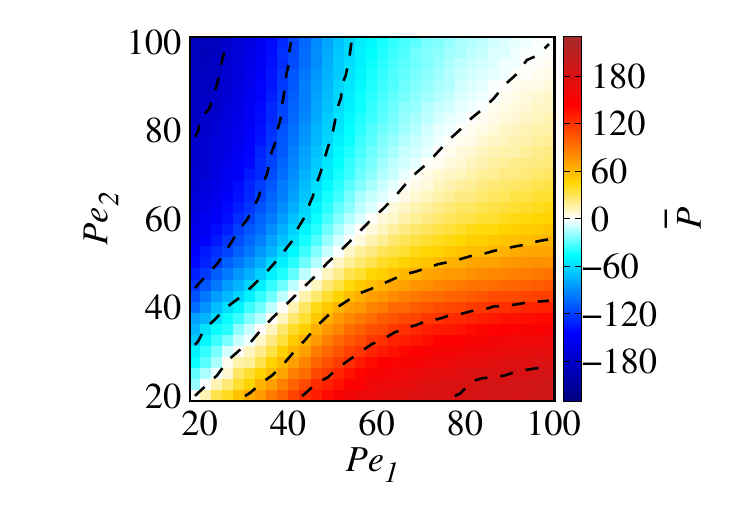} \vskip-2mm (a)
		\end{center}\end{minipage}\hskip-1cm
		\begin{minipage}[t]{0.37\textwidth}\begin{center}
				\includegraphics[width=\textwidth]{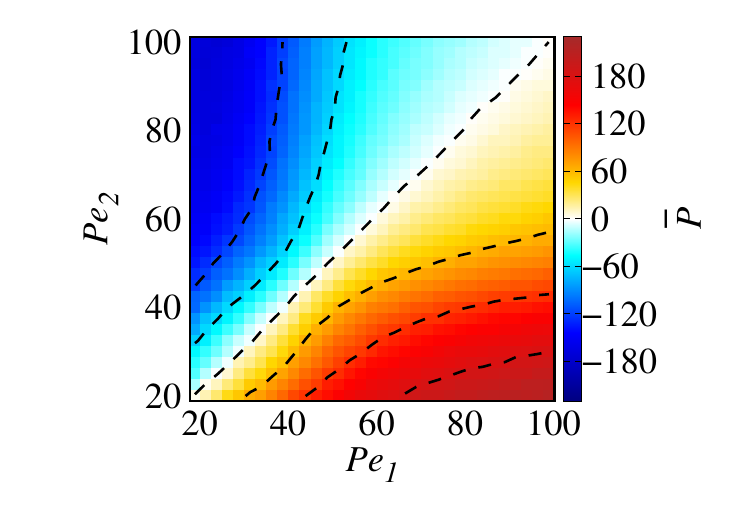} \vskip-2mm (b)
		\end{center}\end{minipage}\hskip-1cm
		\begin{minipage}[t]{0.37\textwidth}\begin{center}
				\includegraphics[width=\textwidth]{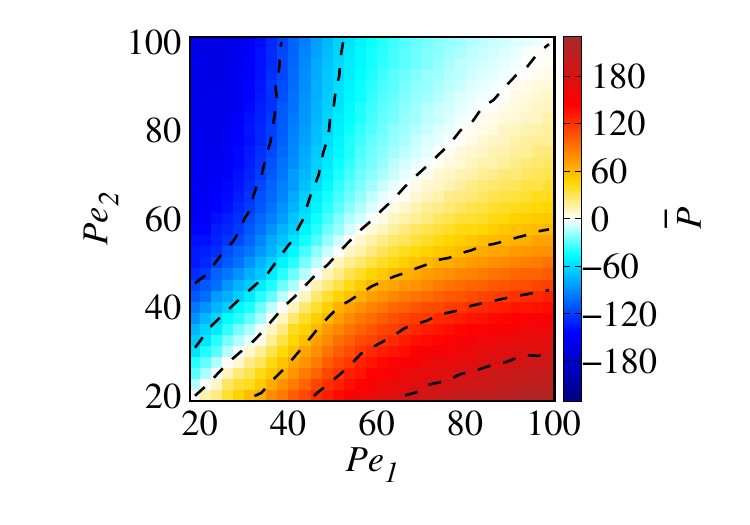} \vskip-2mm (c)
		\end{center}\end{minipage}
		\\
		\begin{minipage}[t]{0.37\textwidth}\begin{center}
				\includegraphics[width=\textwidth]{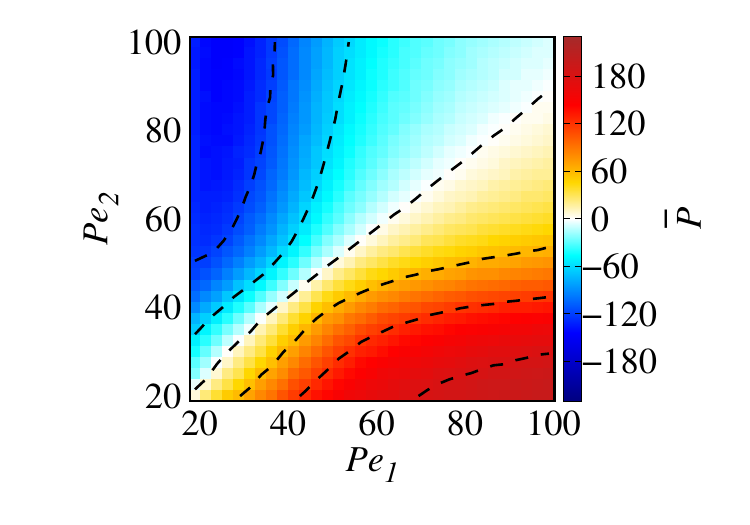} \vskip-2mm (d)
		\end{center}\end{minipage}\hskip-1cm
		\begin{minipage}[t]{0.37\textwidth}\begin{center}
				\includegraphics[width=\textwidth]{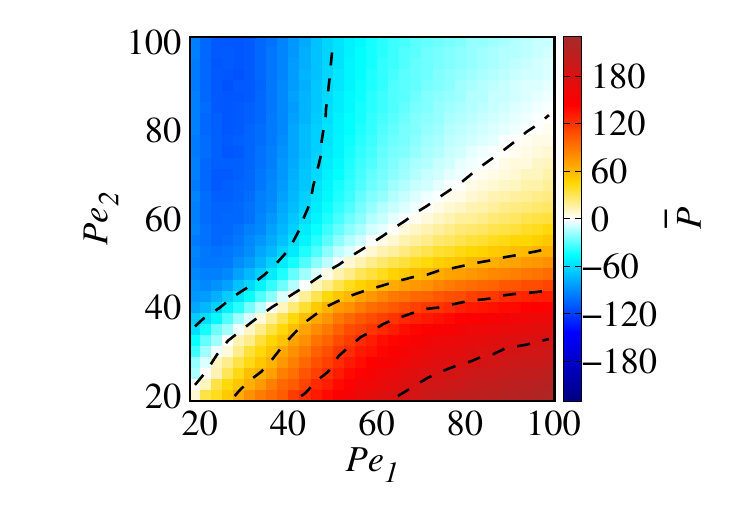} \vskip-2mm (e)
		\end{center}\end{minipage}\hskip-1cm
		\begin{minipage}[t]{0.37\textwidth}\begin{center}
				\includegraphics[width=\textwidth]{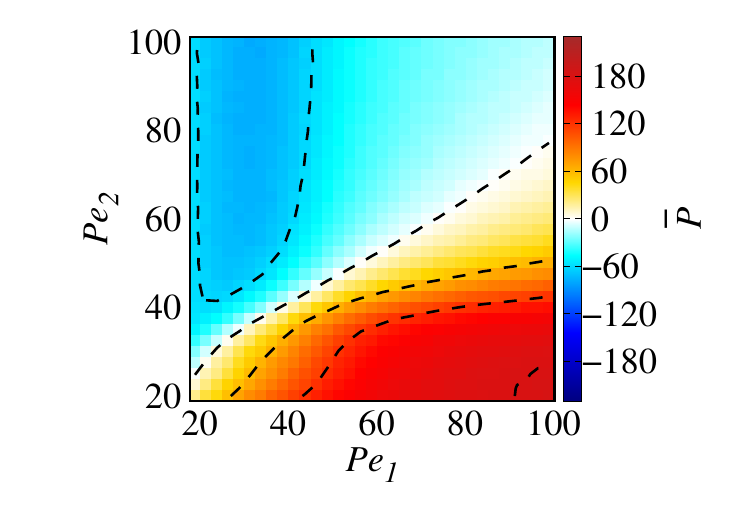} \vskip-2mm (f)
		\end{center}\end{minipage}
		\caption{Rescaled effective pressure acting on a model inclusion as a function of Péclet numbers $\{Pe_1, Pe_2\}$ for interacting active particles with rescaled global average density, $\bar{\rho}=0.128$ ($\phi=0.1$) and enclosing membrane hardness $Pe_p=52.6$. The first row of panels displays the effective pressure on a  rectangular inclusion with different rescaled widths: (a) $L_2/\sigma=20$, (b) $10$, and (c) $5$. The second row of panels shows the effective pressure on a disklike inclusion with different rescaled diameters: (d) $\sigma_c/\sigma=20$, (e) $10$, and (f) $5$. Dashed lines represent pressure contours corresponding to the tick values indicated on the color sidebar for each panel.}
		\label{Rec}
	\end{figure*}
	
	In Sec. \ref{sec:NI}, we analyzed the behavior of noninteracting active particles, and now we will explore the interplay between effective pressure and steric interactions between active particles. To accomplish this, we will enable steric interactions and compare the results for both cases of interacting and noninteracting active particles. In the case of interacting particles, many-body effects come into play. Figure \ref{Rec} displays the effective pressure on the inclusion exerted by interacting active particles for both rectangular (top-row panels) and disklike inclusions (bottom-row panels). By comparing the effective pressure in the case of interacting (Fig. \ref{Rec}) and noninteracting particles (Fig. \ref{NRec}), we observe that the effective pressure in the former case is weaker and varies more gradually with the motility strengths, $Pe_1$ and $Pe_2$. This is due to the weaker accumulation of interacting active particles against the membrane in both regions and a lower magnitude of density difference (see Figs. \ref{den} and \ref{xrden} and compare their bottom-row panels with the top-row panels).
	
	Figures \ref{Rec}(d)-\ref{Rec}(f) present the effective pressure on the disklike inclusion with diameters $\sigma_c/\sigma=20, 10$, and $5$, respectively, in the case of interacting active particles. As the diameter of the inclusion decreases, the absolute value of negative pressures consistently decreases due to the reduction in exterior average density of active particles, resulting in a decrease in the absolute value of negative density difference. Positive pressures increase when the diameter $\sigma_c$ is changed from $20\sigma$ to $10\sigma$ but they decrease the diameter is changed from $10\sigma$ to $5\sigma$. When the motility strength inside the inclusion is significantly less than outside, active particles predominantly or entirely occupy the region inside the inclusion. By decreasing $\sigma_c$, the average density of active particles inside the inclusion increases up to a maximum value corresponding to the density of particles in a lattice with a close-packing hexagonal arrangement, $\bar{\rho}_{hcp}=1.159$. In reality, excluded-volume interactions between active particles prevent the density from exceeding $\bar{\rho}_{hcp}=1.159$. Conversely, the average density of active particles outside the inclusion decreases by reducing the inclusion diameter until it reaches a minimum area capable of trapping all active particles within the inclusion (when the interior average density reaches $\bar{\rho}_{hcp}=1.159$). For this range of $\sigma_c$ values, positive effective pressures increase by reducing the diameter. After that, while the interior average density remains constant, the exterior average density increases due to more active particles remaining outside the inclusion. As a result, the values of positive pressures decrease. In the range of motility strengths ($20\leq Pe_{1,2}\leq 100$) that we consider, the interior average density cannot reach $\bar{\rho}_{hcp}$ for $\sigma_c/\sigma=10$ and $5$. However, when the density difference is maximum for both diameters at $Pe_1=100$ and $Pe_2=20$, the exterior average density of active particles for $\sigma_c/\sigma=5$ is greater than for $\sigma_c/\sigma=10$. As a result, the exterior active pressure weakens the effective pressure more for $\sigma_c/\sigma=5$, resulting in a lower maximum effective pressure than the maximum pressure for $\sigma_c/\sigma=10$. Therefore, the disk with a diameter of $\sigma_c/\sigma=10$ has the largest maximum effective pressure among inclusions with diameters $\sigma_c/\sigma=20, 10$, and $5$. This is different from the case of noninteracting active particles, where the inclusion with a diameter of $\sigma_c/\sigma=5$ has the largest maximum effective pressure.
	
\begin{figure}[t!]
		\begin{center}
			\includegraphics[width=1\textwidth]{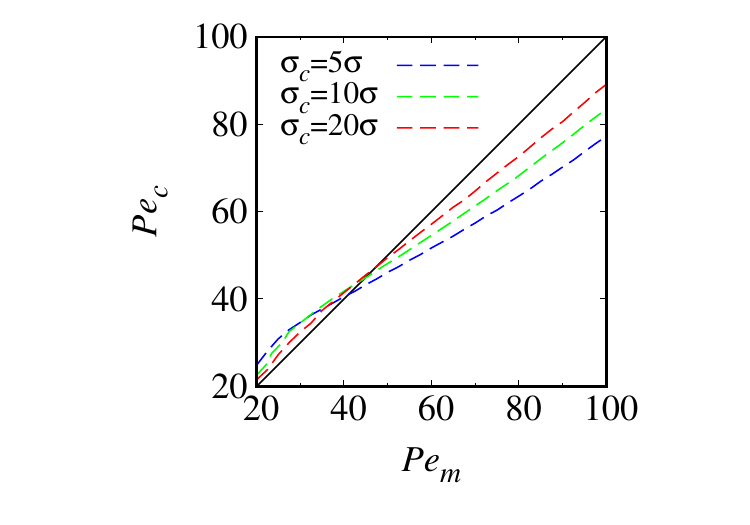}
			\caption{Contour lines of $\bar{P}=0$ as a function of $Pe_1$ and $Pe_2$ for disklike inclusions of different diameters as indicated on the plot. Solid line represents the reference line $Pe_2=Pe_1$.}
			\label{cl0}
		\end{center}
	\end{figure}

	Fig. \ref{cl0} displays contour lines of zero effective pressure for disklike inclusions with different diameters. As the diameter decreases, the contour lines deviate more from the reference line (the line of $Pe_1=Pe_2$). The contour lines lie above the reference line for $20\leq Pe_{1,2}\leq 40$ and below the reference line for $Pe_{1,2}>40$. This is different from the noninteracting case where the contour lines are completely below the reference line (compare Fig. \ref{cl0} and \ref{ncl0}). These contour lines match their contour lines of zero density difference in the range of low and intermediate motility strengths. For high motility strengths, the contour lines of zero pressure lie below their contour lines of zero density difference (data not shown) due to a moderately larger exterior active pressure, causing the effective pressure becoming negative when the density difference is equal to or slightly greater than zero.
	
%

		
	\begin{figure*}[t!]\centering
		\hskip-7mm
		\\
		\begin{minipage}[t]{0.37\textwidth}\begin{center}
				\includegraphics[width=\textwidth]{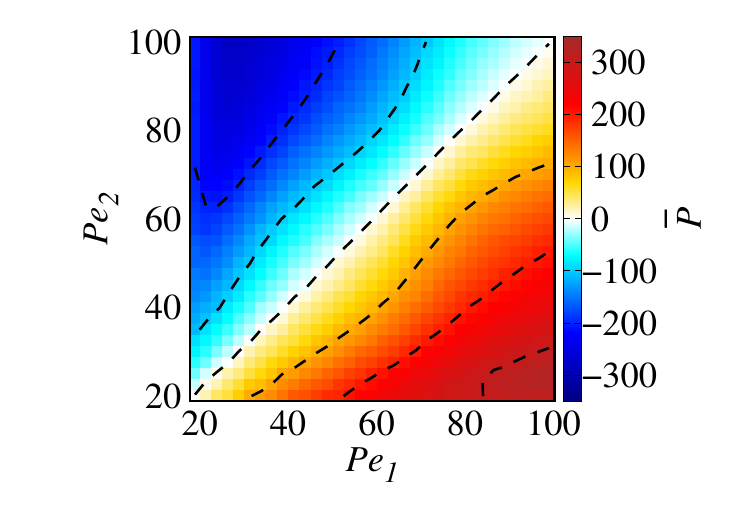} \vskip-2mm (a)
		\end{center}\end{minipage}\hskip-1cm
		\begin{minipage}[t]{0.37\textwidth}\begin{center}
				\includegraphics[width=\textwidth]{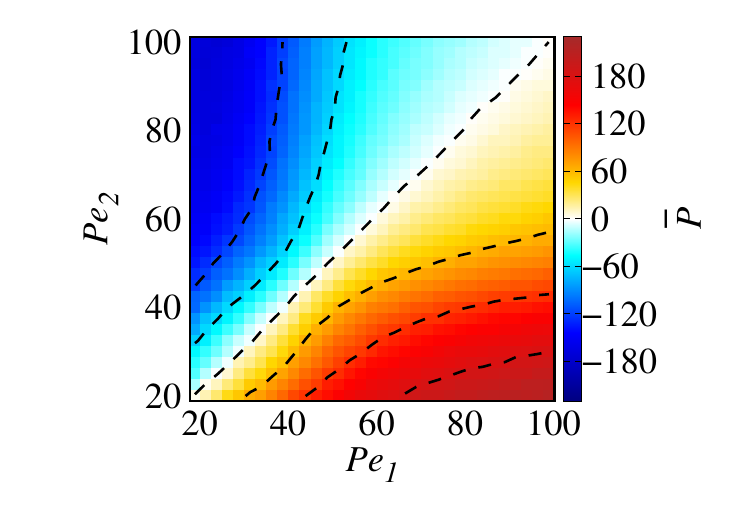} \vskip-2mm (b)
		\end{center}\end{minipage}\hskip-1cm
		\begin{minipage}[t]{0.37\textwidth}\begin{center}
				\includegraphics[width=\textwidth]{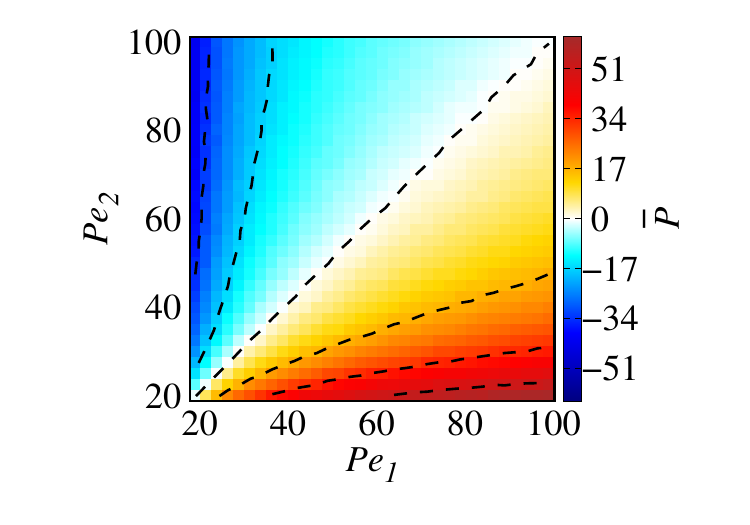} \vskip-2mm (c) 
		\end{center}\end{minipage}
		\\
		\begin{minipage}[t]{0.37\textwidth}\begin{center}
				\includegraphics[width=\textwidth]{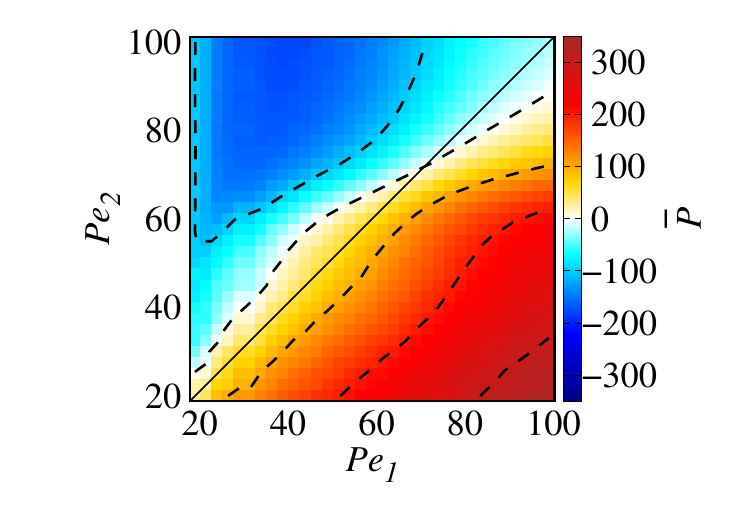} \vskip-2mm (d)
		\end{center}\end{minipage}\hskip-1cm
		\begin{minipage}[t]{0.37\textwidth}\begin{center}
				\includegraphics[width=\textwidth]{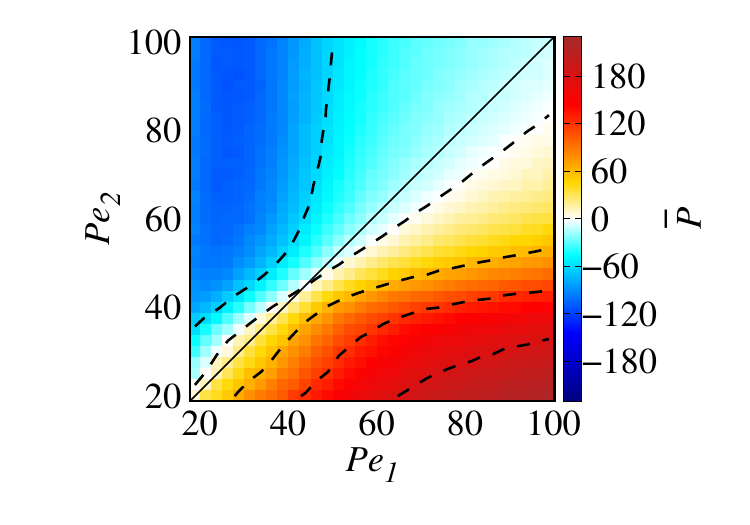} \vskip-2mm (e)
		\end{center}\end{minipage}\hskip-1cm
		\begin{minipage}[t]{0.37\textwidth}\begin{center}
				\includegraphics[width=\textwidth]{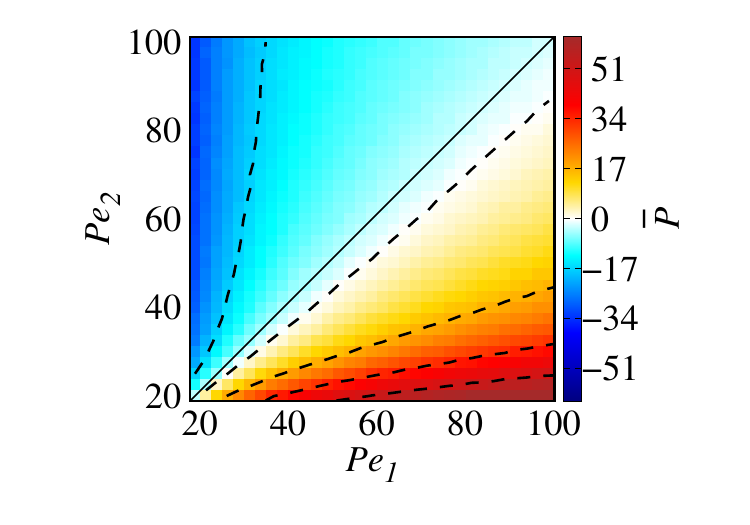} \vskip-2mm (f) 
		\end{center}\end{minipage}		
		\caption{Rescaled effective pressure on the inclusion as a function of Péclet numbers $\{Pe_1, Pe_2\}$ for interacting active particles with a rescaled global average density of $\bar{\rho}=0.128$ ($\phi=0.1$). The first row of panels displays the effective pressure on a rectangular inclusion with a rescaled width of $L_2/\sigma=10$ and different enclosing membrane hardness: (a) $Pe_p=78.9$, (b) $52.6$, and (c) $26.3$. The second row of panels shows the effective pressure on a disklike inclusion with a rescaled diameter of $\sigma_c/\sigma=10$ and different enclosing membrane hardness: (d) $Pe_p=78.9$, (e) $52.6$, and (f) $26.3$. Dashed lines represent pressure contours at the tick values of the color sidebar for each panel. The solid diagonal solid lines depict the reference bisector $Pe_1=Pe_2$.}
		\label{stiff:rec}
	\end{figure*}
		
	\subsection{Role of membrane hardness}
	\label{sec:stiff}

	In this section, we discuss how the hardness of the enclosing membrane affects the effective pressure on the inclusion. Figures \ref{stiff:rec}(a)-\ref{stiff:rec}(c) displays the effective pressure on the rectangular inclusion as a function of motility field strengths, $Pe_1$ and $Pe_2$, for a fixed width of the rectangle, $L_2=10\sigma$, and varying values of the membrane hardness parameter, $Pe_p=78.9$ (a), $52.6$ (b), and $26.3$ (c). The dynamics of the system significantly slows down with increasing $Pe_p$. Thus, to ensure that the effective  pressure for $Pe_p=78.9$  is computed in the steady state, we allow a relaxation time ($3\times 10^7$ time steps) which is an order of magnitude longer than what is used for $Pe_p=52.6$ and $26.3$ at low motility strengths  ($20\leq Pe_{1,2}<30$). As shown, the overall magnitude of the effective pressure increases with the strength of the membrane hardness, $Pe_p$. The areas of larger (lower) effective pressure in the parameter space $\{Pe_1,Pe_2\}$ expand as the membrane hardness increases. This is evidenced by the darker red and blue areas in the panels of Fig. \ref{stiff:rec}. In fact, at specified values of motility strengths, by increasing the strength of the membrane hardness, active particles encounter more resistance when passing through the membrane in the region of lower motility. This results in a stronger average density of particles in that area and intensifies the density difference. Due to the amplified density difference and stronger interaction between active particles and the inclusion membrane, the magnitude of the effective pressure increases with membrane hardness.
	
	For the disklike inclusion, the dependence of effective pressure on the hardness of the membrane is almost similar to the case of the rectangular inclusion. Panels (d-f) in Fig. \ref{stiff:rec} illustrate the effective pressure on the inclusion of diameter, $\sigma_c=10\sigma$, for $Pe_p=78.9$, $52.6$, and $26.3$, respectively. By comparing the contour lines associated with zero effective pressure in these panels, we observe that the range of motility strengths where the contour line lies above the reference line becomes shorter as the membrane hardness decreases. For $Pe_p=26.2$, where the disk is completely soft for this range of motility strengths, we see that the contour line is completely below the reference line.

	\begin{figure*}
		\centering
		\begin{minipage}[b]{0.37\textwidth}
			\centering
			\includegraphics[width=\textwidth]{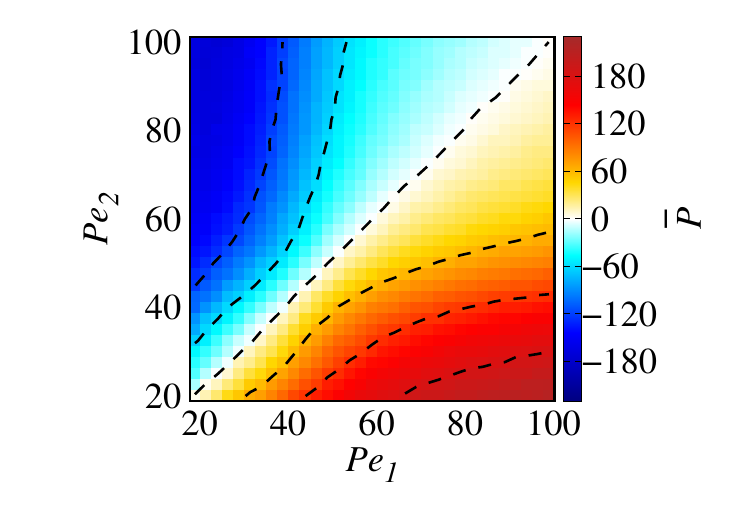}  \vskip-2mm (a)
		\end{minipage}
		\hskip-1cm
		\begin{minipage}[b]{0.37\textwidth}  
			\centering 
			\includegraphics[width=\textwidth]{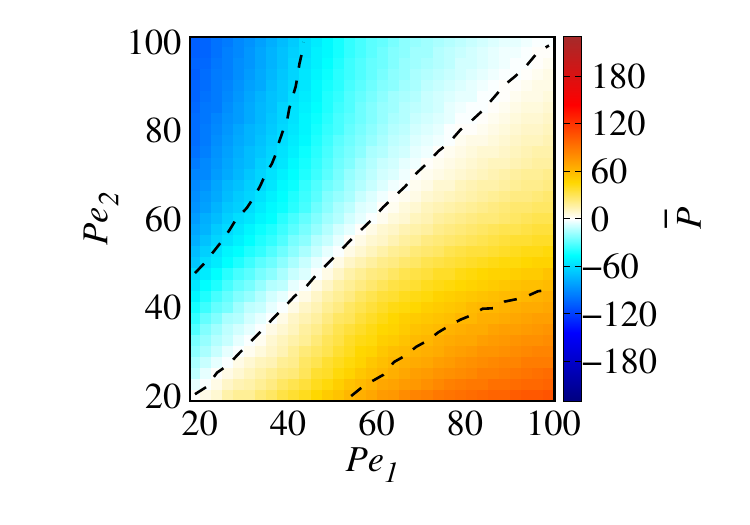} \vskip-2mm (b)
		\end{minipage}
		\vskip\baselineskip
		\begin{minipage}[b]{0.37\textwidth}   
			\centering 
			\includegraphics[width=\textwidth]{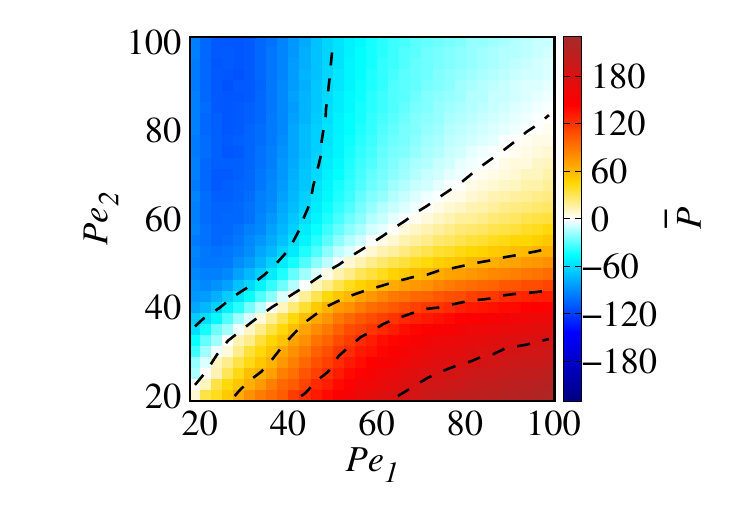} \vskip-2mm (c)
		\end{minipage}
		\hskip-1cm
		\begin{minipage}[b]{0.37\textwidth}   
			\centering 
			\includegraphics[width=\textwidth]{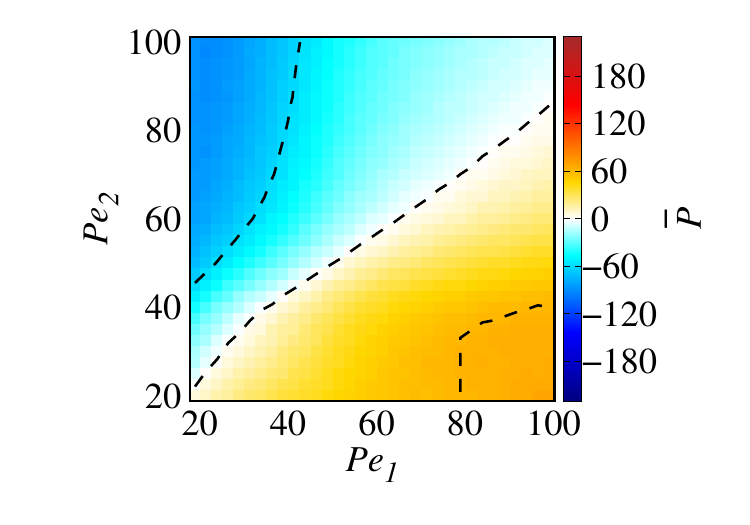} \vskip-2mm (d)
		\end{minipage}
		\caption{Rescaled effective pressure on the inclusion as a function of Péclet numbers ${Pe_1, Pe_2}$ for interacting active particles and enclosing membrane hardness $Pe_p=52.6$. The top row of panels illustrates the effective pressure on a rectangular inclusion with a rescaled width of $L_2/\sigma=10$ and varying values of rescaled global average density: (a) $\bar{\rho}=0.128$ ($\phi=0.1$) and (b) $\bar{\rho}=0.255$ ($\phi=0.2$). The bottom row of panels displays the effective pressure on a disklike inclusion with a rescaled diameter of $\sigma_c/\sigma=10$ and different values of rescaled global average density: (c) $\bar{\rho}=0.128$ ($\phi=0.1$) and (d) $\bar{\rho}=0.255$ ($\phi=0.2$). Dashed lines represent pressure contours corresponding to the tick values on the color sidebar for each panel.}
		\label{osmos}
	\end{figure*}
	
	\subsection{Role of area fraction}
		\label{Sec: area fraction}
	Figures \ref{osmos}(a) and \ref{osmos}(b) display the rescaled effective pressure on a rectangular inclusion with fixed width $L_2=10\sigma$ and varying rescaled global average density of active particles $\bar{\rho}=\{0.128, 0.255\}$, corresponding to area fractions for active particles $\phi=\{0.1, 0.2\}$, respectively. The results illustrate that the rescaled effective pressure weakens as the global average density of active particles increases. The rescaled effective pressure is defined by Eq. (\ref{eq:pres}), which measures the pressure exerted on the inclusion divided by the global average density of active particles. In the case of interacting active particles, the magnitude of the rescaled effective pressure decreases as the density increases because the effective pressure grows with the density at a slower rate than the linear relation. Moreover, an increase in the area fraction enhances the overall magnitude of the effective pressure because the difference in the average density of active particles between the interior and exterior regions increases. However, in the case of a rectangular inclusion, the average density of active particles inside the inclusion cannot reach a saturation value, even for $\phi=0.2$, because the area of the inclusion is large enough to encompass all active particles. This is the reason for the increase in the density difference and in the overall magnitude of the effective pressure.
	
	Figures \ref{osmos}(c) and \ref{osmos}(d) show the rescaled effective pressure exerted on a disklike inclusion with diameter $\sigma_c=10\sigma$ for area fractions of active particles $\phi=0.1$ and $0.2$, respectively. In this case, the magnitude of the rescaled effective pressure also decreases as the area fraction (global average density) increases. The absolute value of the negative effective pressure increases as $\phi$ increases from $0.1$ to $0.2$, but this increase is slower than the linear relation with area fraction (global average density), resulting in a decrease in the magnitude of the negative rescaled effective pressure. For the positive pressure, increasing the area fraction from $0.1$ to $0.2$ not only decreases the rescaled pressure but also decreases the effective pressure because the inclusion can trap up $45\%$ of particles within itself at $\phi=0.2$, compared to $90\%$ at $\phi=0.1$. Consequently, the average density outside the inclusion is generally higher for $\phi=0.2$ than for $\phi=0.1$, resulting in lower positive effective pressures for $\phi=0.2$. Overall, increasing the area fraction strengthens the effective pressure if both the regions inside and outside the enclosure have enough capacity to incorporate all active particles. However, this trend comes to an end when the average density inside or outside the inclusion reaches the saturation value, $\bar{\rho}_{hcp}=1.159$.
	\\

	\section{Summary}
	\label{sec:summary}
	In this study, a minimal active Brownian particle model was employed to investigate the distribution of active particles inside and outside permeable inclusions and the effective pressure exerted on the permeable membrane of rectangular and disklike inclusions that are immersed in an active fluid. The motility strength of the particles was allowed to vary inside and outside the inclusion, and the study examined the role of motility field strengths varied in the range $20\leq Pe_{1,2}\leq 100$. The findings revealed that active particles tend to concentrate more strongly inside the fluid region with weaker motility, leading to stronger active pressure in that area. Consequently, the effective (osmotic-like) pressure is directed outward from the fluid region characterized by lower motility.
	
We have also discussed the differences in effective pressure resulting from the inclusion's rectangular and disklike geometries. These differences are primarily reflected in the constant-pressure contour lines plotted across the parameter space. Our findings show that the contour-line of zero pressure is represented by the straight line of $Pe_1=Pe_2$ for the rectangular inclusion. In contrast, for the disklike inclusion, it appears as a curved line that becomes increasingly straight as the diameter of the inclusion increases. 
	
	In this work, we focused on the impact of mismatched motility fields inside and outside permeable inclusions in the range of $20\leq Pe_{1,2}\leq 100$, corresponding to persistence lengths of $13.3\leq l_p/\sigma \leq 66.7$ for active particles. These persistence lengths are comparable to or larger than the size of the inclusions ($L_2/\sigma$ and $\sigma_c/\sigma$ ranging from $5$ to $20$), where the effect of strong accumulation of active particles near the membrane on the effective pressure is pronounced. We verified that the results do not qualitatively change when we increase the size of the inclusion to sizes  beyond the persistence length while keeping other system parameters fixed. However, the magnitude of the (rescaled) effective pressure increases with $L_2$. It is important to note that in the case of $L_2$ (or $\sigma_c$) being greater than $l_p$, the fraction of particles that accumulate near the membrane inside and outside the inclusion with respect to the total particles present within each region decreases. However, due to the increase in the total number of active particles for increased $L_2$ (or $\sigma_c$) at the same rescaled global average density, the number of active particles that interact with the inclusion membrane increases, resulting in a stronger effective pressure. In the case of interacting particles, there is a limitation on the number of particles that can accumulate near the membrane and interact with it due to excluded volume interactions. This limitation results in a weak increase in the magnitude of the effective pressure with $L_2$ (or $\sigma_c$). However, the increase of effective pressure with the size of the inclusion continues until the number of accumulated particles per unit perimeter of the membrane reaches a maximum value allowed by excluded volume interactions. For simplicity and numerical efficiency, we assumed that the model membrane enclosing the interior region of the inclusion is a continuum boundary layer that interacts with active particles through a short-ranged soft repulsive potential [see eq. (\eqref{SWCA})]. We verified that other choices for the soft repulsive interaction do not affect the qualitative aspects of our results, although the effective pressure on the inclusion may quantitatively vary depending on the specific choice of the potential. Future studies may consider the roles of membrane dynamics and flexibility, as well as inclusion (vesicle) mobility, in the present context (i.e., on the effective pressure) using coarse-grained particle-based models such as those considered in related contexts \cite{S.Zhang2010, Daddi2019a, Daddi2019b, Angelan2016, Vutukuri2020,Hagan2021}, where the membrane can be deformed or even destroyed due to the presence of active particles.
	

	Such extensions should bring the model inclusions used here closer to real-life examples of vesicles and soft permeable objects. These examples may be furnished by fluid enclosures such as lipid and polymer vesicles \cite{Rideau2018,Kamat2011}, immiscible and stabilized emulsion drops, and active droplets \cite{Cates2017,Hyman2014,Naji2018}. Other examples of permeable objects include polymersomes (polymer vesicles) that can be designed with controlled permeability \cite{Joseph2017} and polymer stomatocytes obtained from controlled deformation of polymersomes \cite{Wilson2012}, which have recently been used for the selective entrapment of catalytically active platinum nanoparticles. In biological cells, phagocytosis (representing the cell's ability to ingest and internalize foreign particles) has been used for the controlled internalization and cell-assisted assembly of nonactive polystyrene microparticles, as well as crystallites inside fibroblast cells \cite{Kodali2007}. Our work should inspire further studies in such contexts with active particles utilized as phagocytosed components. Active particles have also been used for targeted drug and cargo delivery \cite{Peng2018,Joseph2017,Magdanz2019,Ghosh2020}. In these contexts, soft tissues such as tumors can play the role of permeable objects as they exhibit enhanced permeation to suitably designed active agents, e.g., platinum-sputtered polymersome nanomotors \cite{Peng2018} and magneto-aerotactic bacteria \cite{Felfoul2016,Ghosh2020}.
	
	Our model is also based on a minimal model for active particles that are taken as disklike particles with merely steric interactions. Therefore, effects of more complex shapes of active particles \cite{Dunkel 2014,Graaf 2016,Filion 2016}, particle chirality, and other types of particle-particle interactions such as Vicsek interaction \cite{Naji2021a,Naji2018} remain to be addressed. We have considered relatively low area fractions of active particles; hence, another possible venue to explore later includes higher area fractions where motility-induced phase separation can emerge \cite{Baskaran2013}.\\

	\section{Acknowledgements}
A.N. acknowledge support from the ICTP (Trieste, Italy) through the Associates Programme and from the Simons Foundation through Grant No. 284558FY19. We  thank L. Javidpour for useful comments and acknowledge support from the High Performance Computing Center, IPM. We also thank anonymous referees for useful suggestions that led to significant improvements of the paper. 
 \appendix
\section{Maximum force from the sWCA potential}
\label{app}
\begin{figure}
	\includegraphics[width=0.85\textwidth]{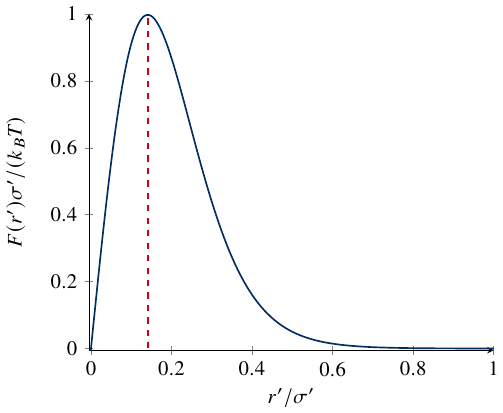}
	\caption{The figure illustrates the rescaled force, $F(r')\sigma'/(k_\text{B}T)$ produced by the sWCA potential is plotted as a function of the rescaled distance from the midline of inclusion membrane [see Eq. (\ref{distance})], $r'/\sigma'$, being varied over the interval $0\leq r'/\sigma'\leq 1$. Here, we have fixed $\tilde{F}_{\mathrm{max}}=1$ and used Eq. (\ref{resF}) to calculate. The red dashed line represents the rescaled distance at which the maximum force occurs ($r'_{\mathrm{max}}/\sigma'=0.141$).}
	
	\label{F}
\end{figure}

The force field $F(r')$ produced by the inclusion membrane can be obtained by differentiating Eq. (\eqref{SWCA}) with respect to ${r'}$ (after replacing  ${r'}_{i}$ with ${r'}$) as
\begin{equation}
		F(r') = 12\tilde{F}_{\mathrm{max}}\,\beta \left[\frac{\sigma'^{\,14}}{({r'}^2+\alpha^{2})^7} -\frac{\sigma'^{\,8}}{2({r'}^2+\alpha^{2})^4}\right]\frac{r'}{\sigma'^2}.
	\label{resF}
\end{equation}

The maximum force occurs at $r'_{\mathrm{max}}=0.141\sigma'$. Figure \ref{F} depicts the rescaled force, Eq. (\ref{resF}), as a function of  $r'/\sigma'$ where $\tilde{F}_{\mathrm{max}}$ is fixed as $\tilde{F}_{\mathrm{max}}=1$. As noted in the text, $\tilde{F}_{\mathrm{max}}$ gives the magnitude of the maximum force produced by the sWCA potential and that $\tilde{F}_{\mathrm{max}} = 2Pe_p$, with  $Pe_p$ being interpreted as the  characteristic P\'eclet number signifying the successful transit of active particles through the inclusion membrane in the absence of thermal noise and interactions between active particles. This can be established as follows. We set the left-hand side of Eq. (\ref{Alang1}) to $0$, ignore the translational thermal noise, $\eta_i(t)$,  and replace the term $-{\partial U}/{\partial{\mathbf r}_i}$ with $\tilde{F}_{\mathrm{max}}k_\text{B}T/\sigma'$. Since the range of the sWCA potential is  equal to the diameter of active particles [$\sigma'=\sigma$, see Eqs. (\ref{w}) and (\ref{beta})], we have
\begin{equation}
	(\mu_{T}k_\text{B}T/\sigma)\tilde{F}_{\mathrm{max}}=v_p,
\end{equation}
where $v_p$ is the minimum velocity magnitude for an active particle to pass through the membrane. Utilizing $\mu_T = D_T/(k_\text{B}T)$ and the corresponding P\'eclet number, $Pe_p=v_p\sigma/(2D_T)$, see Eq. (\ref{Pe}), we then have
\begin{equation}
	\tilde{F}_{\mathrm{max}}= 2Pe_p. 
	\label{Fmax}
\end{equation}

	
\end{document}